\documentclass[%
 reprint,
 showkeys,
superscriptaddress,
amsmath,amssymb,
aps,
floatfix,
]{revtex4-2}

\newcommand{\beginsupplement}{%
        \setcounter{table}{0}
        \renewcommand{\thetable}{S\arabic{table}}%
        \setcounter{equation}{0}
        \renewcommand{\theequation}{S\arabic{equation}}%
        \setcounter{figure}{0}
        \renewcommand{\thefigure}{S\arabic{figure}}%
        }
    
\usepackage{graphicx}
\usepackage{dcolumn}
\usepackage{bm}
\usepackage{textcomp}
\usepackage{color,soul}
\usepackage{lipsum}
\usepackage{parskip} 
\usepackage[normalem]{ulem}	
\usepackage{epstopdf}
\PrependGraphicsExtensions{.tif, .tiff}
\AppendGraphicsExtensions{.tif}

\DeclareMathOperator{\var}{var}

\newcommand{\AR}{{\rm AR}}
\newcommand{\D}{{\rm d}}
\newcommand{\lb}{{\ell_{\rm b}}}

\begin{document}


\title{Trade-offs in phenotypic noise synchronize emergent topology to actively enhance transport in microbial environments} 


\author{Jayabrata Dhar}%
\affiliation{Physics of Living Matter, Department of Physics and Materials Science, University of Luxembourg, 162 A, Avenue de la Faïencerie, L-1511 Luxembourg City, Luxembourg}

\author{Anh L. P. Thai}
\affiliation{Physics of Living Matter, Department of Physics and Materials Science, University of Luxembourg, 162 A, Avenue de la Faïencerie, L-1511 Luxembourg City, Luxembourg}

\author{Arkajyoti Ghoshal}%
\affiliation{Physics of Living Matter, Department of Physics and Materials Science, University of Luxembourg, 162 A, Avenue de la Faïencerie, L-1511 Luxembourg City, Luxembourg} 

\author{Luca Giomi}%
\affiliation{Instituut-Lorentz, Universiteit Leiden, P.O. Box 9506, 2300 RA Lieden, The Netherlands} 

\author{Anupam Sengupta}%
\email{anupam.sengupta@uni.lu}
\affiliation{Physics of Living Matter, Department of Physics and Materials Science, University of Luxembourg, 162 A, Avenue de la Faïencerie, L-1511 Luxembourg City, Luxembourg} 


\begin{abstract}

Phenotypic noise underpins homeostasis and fitness of individual cells. Yet, the extent to which noise shapes cell-to-population properties in microbial active matter remains poorly understood. By quantifying variability in confluent \textit{E.coli} strains, we catalogue noise across different phenotypic traits. The noise, measured over different temperatures serving as proxy for cellular activity, spanned more than two orders of magnitude. The maximum noise was associated with the cell geometry and the critical colony area at the onset of mono-to-multilayer transition (MTMT), while the lower bound was set by the critical time of the MTMT. Our results, supported by a hydrodynamic model, suggest that a trade-off between the noise in the cell geometry and the growth rate can lead to the self-regulation of the MTMT timing. The MTMT cascades synchronous emergence of hydrodynamic fields, actively enhancing the micro-environmental transport. Our results highlight how interplay of phenotypic noise triggers emergent deterministic properties, and reveal the role of multifield topology--of the colony structure and hydrodynamics--to insulate confluent systems from the inherent noise associated with natural cell-environment settings. 
 
\begin{description}
\item[Note]

The manuscript is supported and accompanied by a Supplementary Information section.
\end{description}
\end{abstract}


\maketitle


\section{\label{sec:Intro}Introduction}

Microbial life, at the level of individual cells, is inherently noisy due to cell-to-cell variability. Triggered by the intrinsic stochasticity of gene expressions \cite{Thomas2018,Schmiedel2019, Thomas2019}, alongside biotic and abiotic factors offered by the microbial habitat \cite{Hilfinger2011,Lidstrom2010,Korobkova2004}, variability in phenotypic traits is frequently observed within mono- and polyclonal populations. In bacteria, phenotypic heterogeneity and the resulting cellular physiology can enhance cellular fitness and functionality \cite{Hashimoto2016,Cerulus2016,Patange2018}, adapt chemotactic attributes \cite{Waite2018,Emonet2008}, support homeostasis \cite{Taheri2015,Schwabe2011}, and regulate bet-hedging strategies in face of environmental perturbations \cite{Acar2008,Ackermann2015}. Whether ramifications of cell-to-cell variability on growing populations are suppressed or reinforced have been studied for single phenotypic contexts \cite{Koutsoumanis2013}, yet in nature, phenotypic shifts are known to occur concurrently \cite{Dougherty2014,Thoming2020}. It thus remains to be understood how variability across multiple phenotypic traits co-emerge as biological activity changes, and crucially, if cross-talks therein provide intrinsic mechanistic cues that could lead to deterministic statistical properties of a growing population.


Many bacterial species exhibit surface-association, and have been long studied for their significance in ecology, medicine, and industry. Recent single-cell studies, coupled with mechanistic modeling, have revealed critical roles that single-cell geometry and growth dynamics play in shaping the dynamical properties of growing bacterial layers \cite{You2018,Marenduzzo2018,Echten2020,Warren2019,Beroz2018,Hartmann2019}. The emergence of structural order and low-dimensional topological attributes in expanding colonies, including singularities (topological defects), have been shown to create biomechanical settings that promote morphogenesis of a layered colony (mono- or multilayered bacterial colonies) into well-developed 3D biofilm over much longer timescales. Specifically, the mono-to-multilayer transition (MTMT)--a key step in biofilm initiation--emerges due to a mechanistic interplay of geometry, order and topology \cite{You2019,Sengupta2020}, as has also been observed in the case of motile bacterial swarms \cite{Copenhagen2021,Patteson2018}. The ability of bacteria to collectively exploit topological defects for optimal navigation strategies \cite{Meacock2021}, localizing sporulation sites \cite{Yaman2019}, and potentially, for driving local nutrient fluxes \cite{Mathijssen2018,Lastra2021}, showcases emergent functionalities that bacteria can access over different physiological timescales. 

Despite these recent mechanistic insights, the role of cell-level variability in shaping emergent population-scale traits has remained overlooked. Specifically, if and how phenotypic variability manifests in colony-scale properties and dynamics are poorly understood. More importantly, how activity shapes \textit{noisy} cell-level statistics, and if phenotypic noise could give rise to \textit{deterministic} colony traits, await systematic investigation, particularly in the context of behavioural and physiological functions in microbial active matter.

Here, we combine single-cell time-lapse imaging, particle image velocimetry, numerical simulations and continuum modelling to quantify cell-level variability across key phenotypic traits, and analyze their role in tuning emergent properties of confluent bacterial colonies. Our study spans cell-to-colony development of \textit{E. coli} strains C600-wt and NCM3722 delta-motA (hereon referred to as Strain-1 and Strain-2 respectively), growing independently in nutrient-replete conditions under different temperatures which serve as proxy for biophysical activity. By tracking the cellular traits over multiple generations leading up to the mono-to-multilayer transition (MTMT), we first capture the statistical distributions of phenotypic traits to arrive at trait-specific cellular noise terms. We use the dynamic experiments to extract the properties of confluent colonies at the critical time point, $t_{c}$, the onset time of the MTMT event. Our results, with $t_{c}$ as the reference time point, reveal that cellular noise, spanning over two orders of magnitude, is maximum for the cells' aspect ratio (\AR{}) and the critical colony area ($A_{\rm c}$), while the critical timing of MTMT ($t_{c}$) is least noisy--and thus statistically deterministic--across the strains and temperatures considered in this work. Confluent bacterial colonies represent multifield topological systems \cite{Giomi2017}, wherein the topology of colony structure and that of the hydrodynamic fields co-emerge synchronously, with respect to the critical MTMT time, $t_{c}$. We argue that the deterministic timing of the MTMT event emerges due to the trade-offs between the noise in the aspect ratio (a cell level trait) and the growth rate (a population-scale descriptor), leading to self-regulated critical time of the onset of the MTMT. Our experimental observations are rationalized with the help of a continuum model of isotropically expanding colonies, by which we demonstrate that both the mean and variance of the critical area are unaffected by the temperature, whereas the critical time becomes more and more deterministic as temperature is increased. The MTMT-hydrodynamics synchrony actively enhances microscale transport in the vicinity of the expanding colony. By linking phenotypic noise noise to emergent colony-scale properties, our work establishes a mechanistic basis for minimizing variability at colony-scales, within an otherwise noisy microbe-environment setting.

\begin{figure}[htbp]
\includegraphics[scale=0.13]{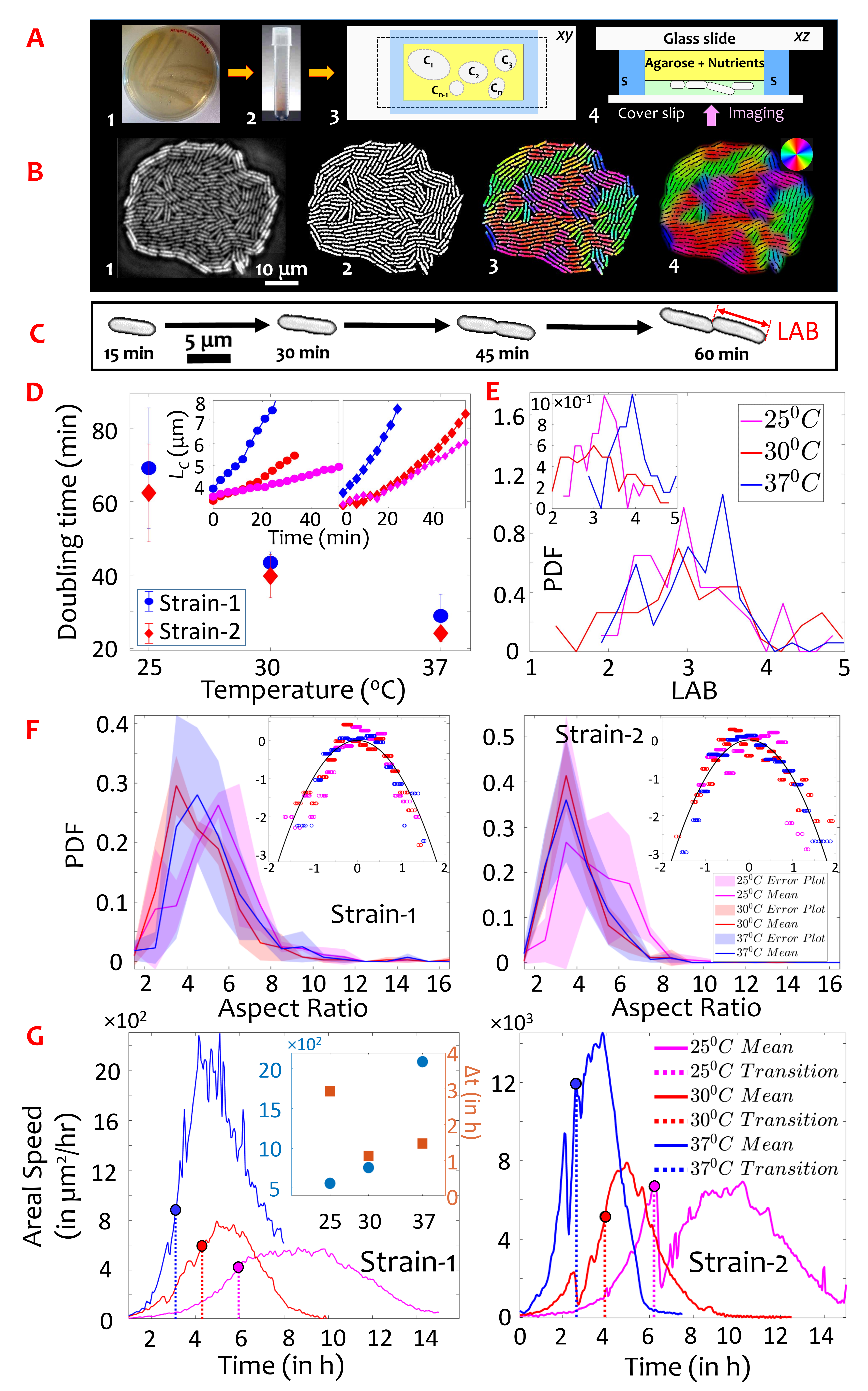} 
\caption{\label{fig:FIG1}\textbf{Activity-dependent phenotypic traits and  microscale dynamics.} 
\textbf{(A)} Experimental steps used in our experiments, starting with plate streaking (on a nutrient-replete agar plate (\textbf{1}); followed by growth of liquid culture (\textbf{2}). (\textbf{3}) and (\textbf{4}) show the bacterial growth pool, used to capture phase-contrast time-lapse microscopy for colony growth.
\textbf{(B)} Phase-contrast raw images of the growing colonies (at least 3 experimental replicates in each case) are and binarized after appropriate thresholding \textbf{(1,2)}, and colour-coded to present the local cell orientations \textbf{(3,4)}. 
\textbf{(C)} Single cell division is captured using time-lapse imaging, the length-at-birth (LAB) is shown in the last panel. \textbf{(D)} The population growth, implicitly related to the cell elongation rate, is as a function of growth temperature. Corresponding doubling times from the area growth and cell number are shown in Supplementary Figure \ref{fig:suppFig1}. \textbf{(E)} The PDF for the LAB for Strain-2 and inset Strain-1. The PDF shows that the LAB has a strong variance across all temperatures. complementing the cell \AR{} data (panel \textbf{(F)}). The \AR{} PDF distribution in the colony just prior to the MTMT. In each plot, the shaded patch denotes the standard deviation error in the observation while the mean is depicted by a solid line. Inset: the quantity $\log\sqrt{2\pi\var\AR{}}\,PDF$ versus $(\log\AR{}-\langle\log\AR{}\rangle )^2/\sqrt{2\var\log\AR{}}$. All data points collapse onto the same parabola indicating that \AR{} is log-normally distributed, regardless the specific temperature and strain. \textbf{(G)} Areal speed, $\D A/\D t$ (in $\mu$m$^{2}$/h), for each temperature and strain links the colony-scale geometry to the biological activity. The corresponding vertical dashed line indicates the point of MTMT. Inset \textbf{(G)} The maximum areal speed (blue) increases, while the interval $\Delta t$ (maroon) between the peak speed and the MTMT shrinks as temperature goes up.
}
\end{figure}

\section{\label{sec:Results}Results}

\subsection{\label{sec:ActivityProxy} Growth temperature is a proxy for activity}

We use growth temperature, a key determinant of physiology, viability and pathogenicity of bacteria \cite{Shehata1975,Trueba1982,Kumar2013,Noll2020}, to regulate the biophysical activity of proliferating colonies. The effect of temperature on bacterial metabolism are well supported by mathematical models and data-backed empirical formulations \cite{Noll2020}, yet temperature-dependent tuning of the biophysical activity, at either individual or colony-scales, has remained unexplored in bacterial active matter. Specifically, by varying the temperature over ecologically relevant values: 25$^{\circ}$C, 30$^{\circ}$C, and 37$^{\circ}$C (see Supplementary Material Section \ref{sec:ExpSetup}), the growth dynamics were controlled at the colony-scale, quantified by the population growth rate, $\D N/\D t$, and by the areal speed, $\D A/\D t$, where $N$ and $A$ denote the number of cells and the colony area at a given time $t$. At an individual-scale, the biophysical activity was measured by the cell elongation rate, $\D \ell/\D t$, $\ell$ being the instantaneous length of an individual cell. Effectively, the choice of growth temperature as a tunable parameter allowed us to regulate the active mechanics underpinning expansion of bacterial colonies over multiple generations. 

Imaging and quantitative tracking of the individual cells across multiple generations, as shown in Figure \ref{fig:FIG1}A-C (see Supplementary Material, Section \ref{sec:TL} and \ref{sec:ImageAnalysis} for the tracking methods), revealed that the doubling time, $\tau_{\rm d}$, the time taken for the population to double in cell number, decreased from 56.73 $\pm$ 2.28 min (44.9 $\pm$ 2.6 min) for the Strain-1 (Strain-2) at 25$^{\circ}$C to 24 $\pm$ 1.12 min (18.73 $\pm$ 0.85 min) at 37$^{\circ}$C (Figure \ref{fig:FIG1}D and Figure \ref{fig:suppFig1}A,B). Corresponding growth rate in terms of change of cell numbers ($\D N/\D t$) exhibited a similar trend with increasing growth temperature for both the {\em E. coli} strains (Figure \ref{fig:suppFig1}C). Thus, at the scale of the bacterial colony, the biophysical activity could be tuned systematically, with the activity at 37$^{\circ}$C stronger by factor of $\approx$3 relative to that at 25$^{\circ}$C.  

Commensurating the colony-scale activity, the elongation of cells at individual scale showed comparable variation (inset, Figure \ref{fig:FIG1}D and Figure \ref{fig:suppFig1}C). The individual cells elongated with a doubling of 28.87 $\pm$ 5.9 min (24.142 $\pm$ 2.26 min) for Strain-1 (Strain-2) at 37$^{\circ}$C, increasing up to 43.39 $\pm$ 2.93 min (39.72 $\pm$ 5.9 min) at 30$^{\circ}$C and to 69.13 $\pm$ 16.4 min (62.38 $\pm$ 13.3 min) 25$^{\circ}$C respectively. Taken together, the growth temperature provided an effective means to tune the biophysical activity in expanding bacterial colonies, at both individual and population scales. 

\subsection{\label{sec:ActivityTopo} Activity shapes phenotypes of cells and colonies}

The temperature-induced modulation of the bacterial physiology manifests in the phenotypes of the individual cells. Figure \ref{fig:FIG1}E shows the distribution of the cell length at birth (LAB), $\lb$, for the three different temperatures. In agreement with previous reports \cite{Campos:2014,Amir:2014}, our results reveal the broad distribution (high variance) in $\lb$ across all temperatures and strains considered here. We tracked the growth of individual cells, alongside colony size, using time-lapse imaging at a regular interval of 3 minutes (see Figure \ref{fig:suppFig2}). Each imaging experiment lasted for 15 h to 18 h, allowing us to capture the mono- to bilayer transition (MTMT, Figure \ref{fig:suppFig3}) and further transitions from bi- to tri- and quadri-layers. Overall, at the MTMT, the cell length at birth, $\lb$ showed a weak dependence on the growth temperature, with broad distributions that peaked within the range of 3 $\mu$m -- 3.9 $\mu$m for Strain-1 and between 2.9 $\mu$m and 3.5 $\mu$m for Strain-2 (inset and Figure \ref{fig:FIG1}E respectively). Though the peaks of $\lb$ at 25$^{\circ}$C and 30$^{\circ}$C differ conspicuously relative the peak value at 37$^{\circ}$C, no detectable difference could be captured between the peaks at 25$^{\circ}$C and 30$^{\circ}$C.

The distribution of the cell aspect ratio, \AR{}, capturing the ratio between the cell length and the width, shows high variance across the temperatures. Figure \ref{fig:FIG1}F shows the \AR{} probability density function (PDF), just prior to the MTMT. For both strains, these PDFs do not reveal specific trends as the temperature is varied within the range 25\textdegree{}C $\le T \le$ 37\textdegree{}C. By contrast, the quantity $\log \AR{}$ exhibits a prominent Gaussian distribution (insets in Figure \ref{fig:FIG1}F), thus indicating that the cellular length prior to the MTMT is log-normally distributed. Furthermore, prior to the MTMT, \AR{} is uniform across the colony and decreases in average with time while maintaining a log-normal distribution of the \AR{} across generations. Interestingly, although no spatial gradient in cell geometry (from the colony center to the periphery) was observed, our results indicate that the mean \AR{} drops with age of the colony. This makes a confluent system additionally prone to the MTMT as the colony grows, since the cells get smaller as the generation number increases. Consequently, at the scale of the colony, the number packing fraction ($i.e.$, the number of cells within a given area) goes up as the colony grows, leading to more efficient packing of cells within the available area. The high growth-induced stresses at the colony center \cite{You2019}, combined with the reduction of mean \AR{} and increment of the number packing fraction, localizes the MTMT closer to the center of the bacterial colony, than in the vicinity of the colony periphery.

The log-normal distribution of the cell \AR{} (Figure \ref{fig:FIG1}F) is frequently observed in a variety of experimental settings trend is a consequence of the random distribution of variables, and in nature, this can be observed frequently in a range of settings \cite{Campos:2014,Amir:2014,Loffler2017eLife,Hosoda2011PRE} and, consistently with the theoretical literature \cite{Amir:2014}, is reflected in the distribution of the LAB (Figure \ref{fig:FIG1}E), signifying a qualitative mapping between the two phenotypic traits considered here. Theoretically, as we show in Section \ref{sec:ActivityHydTransModel} (Equation \ref{eq:mu_sigma}a), the probability distribution of the \AR{} is ultimately determined by that of the LAB, as long as the strain-specific variability in growth rate remains low, validating previous experimental reports \cite{WALLDEN2016Cell}. 


The colony expansion rate, quantified as $\D A/\D t$, varied systematically with the growth temperature (Figure \ref{fig:FIG1}G). Irrespective of the strain and temperature, the areal speed increased monotonically with time, reaching a maximum after the MTMT event synchronously, ultimately dipping to $\sim$0 $\mu$m/s. Higher the growth temperature ($i.e.$, biophysical activity), faster was the maximum areal speed reached. As shown in Figure \ref{fig:FIG1}G (left panel), the maximum areal speed decreased 4-fold for Strain-1, from 2284 $\mu$m$^{2}$/h to 572 $\mu$m$^{2}$/h, as temperature dropped from 37$^\circ$C to 25$^\circ$C. Similarly, for the Strain-2 (Figure \ref{fig:FIG1}G, right panel), the reduction was 2-fold, from 1455 $\mu$m$^{2}$/h to 685 $\mu$m$^{2}$/h. Overall, enhanced biophysical activity accelerates the sequence of events that occur in an expanding colony, as evidenced by the shortening of the time interval between the MTMT and the $(dA/dt)_{max}$ (inset, Figure \ref{fig:FIG1}G): the time interval increases from 175 min. at 37$^{\circ}$C to 88 min. at 25$^{\circ}$C for the Strain-1.


It is worth noting here that the temperature-dependent rates of colony and cell length expansions, at MTMT, trigger differences in the domain formation pattern, and consequently, the ensuing topological defect number and locations \cite{You2018}. Our results show that at the MTMT onset, the number of defects remains same across all temperatures (for each strain, Figure \ref{fig:suppFig4ab}, panel C). However, since $t_{c}$ $\sim T^{-1}$, the time-averaged rate of defect formation at the onset of MTMT event goes up with the temperature (Figure \ref{fig:suppFig4ab}, panel D). Microdomain formation \cite{You2018} and evolution of defects \cite{Sengupta2020} are not the focus of the present study, however the two closely mediate the onset of the MTMT event \cite{You2018,Beroz2018}. In the following, we discuss the implications of phenotypic variabilities on the topological attributes--structural and hydrodynamic--inherent to expanding confluent bacterial colonies.

\begin{figure*}
\includegraphics[scale=0.16]{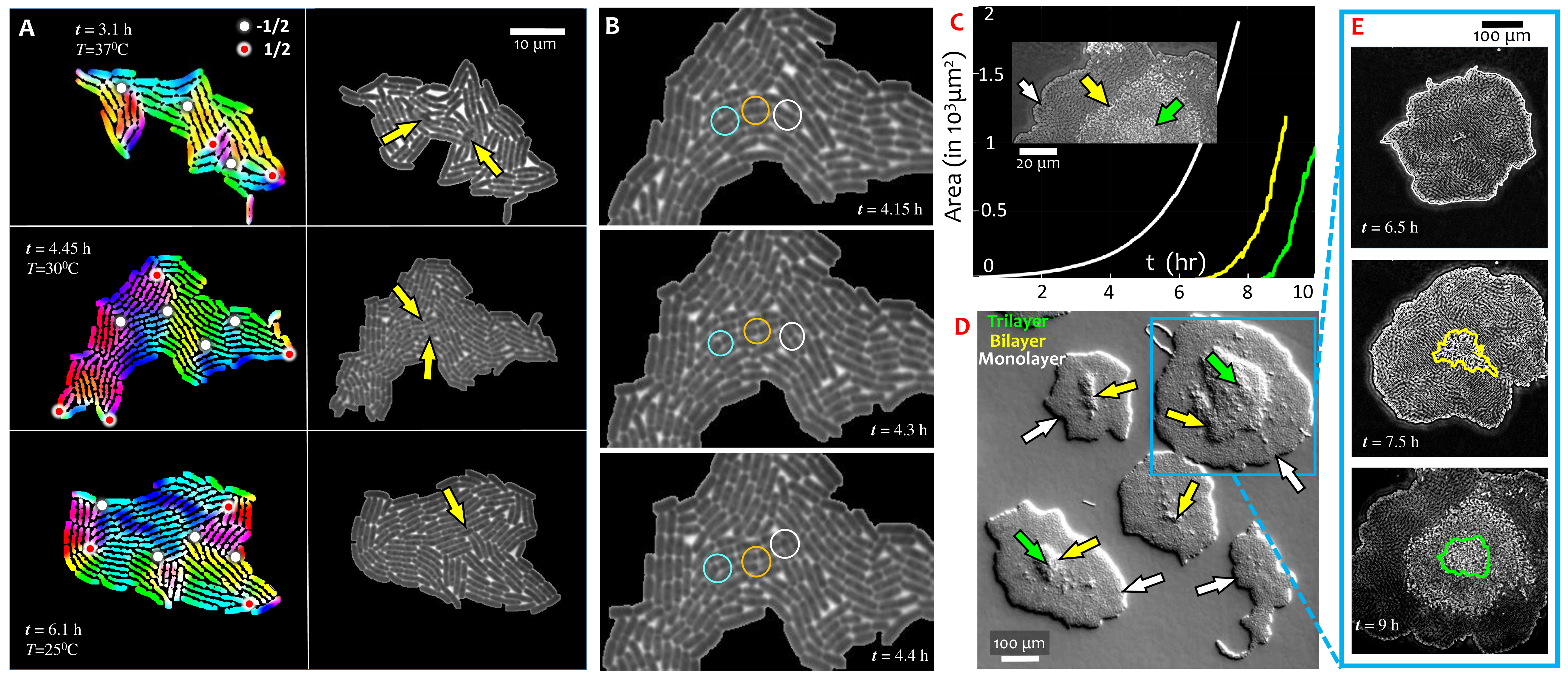} 
\caption{\label{fig:FIG2}\textbf{Defect localization and kinetics at the mono- to multi- layer transition.} 
\textbf{(A)} Topological defects embedded within bacterial colonies  at three different growth temperatures. The images displayed are just before (left column) and at the point of MTMT (right column). For a given strain, the number of topological defects at MTMT remains similar across all temperatures, however the growth rate of defects increases with temperature (Supplementary Figure \ref{fig:suppFig4ab}). \textbf{(B)} Cell division leads to MTMT (representative image for 30\textdegree{}C). The sequential image frames (from top to bottom)  capture the monolayer structure (top), the instance of cell division (middle) and the triggering of MTMT (bottom), visualizing that MTMT follows cell division event as smaller cells are easily extruded out from the surface layer. Multiple MTMT events can simultaneously occur, as indicated by the cyan, red and white circles. \textbf{(C)} The growth of the mono-, bi- and tri- layers show different expansion rates, captured in a $A(t)$ plot, following the colour scheme in panel (D). \textbf{(D)} Overview of multilayer bacterial colonies, shown here are mono- (white), bi- (yellow), and tri-layer (green) structures highlighted with correspondingly coloured arrows. The inset in \textbf{(C)} shows the zoomed-in micrograph of all three layers, the entire colony and expanding layers are shown in panel \textbf{(E)}.
}
\end{figure*}

\subsection{\label{sec:ActivityVerti} Cell geometry determines the kinetics of topological defects}

Active topological manifestations in expanding bacterial colonies \cite{You2018, You2019, Sengupta2020,Marenduzzo2018,Yaman2019}, are due inherently to the anisotropy of bacterial cell shape. Topological defects nucleating at the intersection of nematic microdomains, together with the anisotropic stresses, underpin the biomechanics of expanding bacterial layers. We describe the role of topological attributes close to MTMT, specifically in relation to systematic variation of growth-mediated active stresses. Figure \ref{fig:FIG2}A (left column) enumerates the topological defects accompanying the nematic microdomains, shown here in false colours that represent the local orientations \cite{You2018}. As the biophysical activity is tuned via the growth temperature, the number of defects at the MTMT, $N_{d}$ and the defect concentration ($C_{d}=N_{d}/A$ ($A$ being the colony area), does not vary significantly, remaining nearly uniform for both strains of bacteria (Supplementary Figure \ref{fig:suppFig4ab}). For Strain-1, the average $N_{d}$ ranged between 6 $<N_{d}<$ 8, while for Strain-2, 8 $<N_{d}<$ 11. At 37$^{\circ}$C, 30$^{\circ}$C, and 25$^{\circ}$C, the number of defects for Strain-1 (Strain-2) varied respectively as: 6 $\pm$ 1 (8 $\pm$ 1), 7 $\pm$ 2 (11 $\pm$ 1), and 5 $\pm$ 2 (10 $\pm$ 3). In contrast to $N_{d}$, the rate of generation of defects, $\Delta N_{d}/\Delta t$, exhibited a clear dependence on the biophysical activity, increasing monotonically with temperature. The dependence of $\Delta N_{d}/\Delta t$ on growth temperature (Figure \ref{fig:suppFig4ab}) can be explained by the time elapsed till MTMT: stronger the biophysical activity, shorter is the time taken to reach MTMT, hence higher is the $\Delta N_{d}/\Delta t$. 

The cell aspect ratio (\AR{}), in contrast to biophysical activity, plays a more profound role in regulating the $N_{d}$. Corresponding to the peak \AR{}, 4.5, 3.5, and 5.5 (Figure \ref{fig:FIG1}F, for Strain-1), the mean $N_{d}$ was measured to be 6, 7, and 5 respectively, in the order of decreasing growth temperatures. This reveals a negative correlation of $N_{d}$ with the peak \AR{} at MTMT, such that overall, a lower mean \AR{} corresponds to higher $N_{d}$, and vice-versa. For a given colony area, cells with lower \AR{} self-organize into larger number of nematic microdomains, each with smaller areas \cite{You2018}. Consequently, colonies made of small \AR{} cells embed larger number of potential sites where topological defects can nucleate ($i.e.$, at the intersection of three or more nematic microdomains). This is further confirmed by the inter-strain differences captured across all temperatures in our experiments: The Strain-1, with mean $\AR \sim 5.26$, furnishes a lower range of 6 $<N_{d}<$ 8 relative to the Strain-2, mean $AR \sim 4.1$, that shows higher range 8 $<N_{d}<$ 11 (Supplementary Figure \ref{fig:suppFig4ab}). Since topological defects are not expected for $AR \sim 1$ (spherical cell), we conclude that the $N_{d}$ first increases with 1 $<AR<$ 4 and then goes down as \AR{} increases further. Taken together, our results demonstrate that the kinetics of topological defects in an expanding colony is regulated, weakly, by the temperature-dependent biophysical activity; and more profoundly, by the cell aspect ratio. Thus, in a dynamic setting, the defect kinetics is set by the environment-mediated cell aspect ratio, which can thus be considered as a fundamental determinant of the number and generation rate of topological defects. 

The transition from a mono- to a multilayer structure is triggered by freshly divided cells within the P-zone, a region within the confluent colony which supports MTMT, due to favourable competition between the growth-induced in-plane active stresses and the surface-induced vertical restoring forces \cite{You2019}. Thus, surface properties, along with cell aspect ratio play a key role in MTMT \cite{Beroz2018,You2019}. Figure \ref{fig:FIG2}A (right column) indicates the transition points, shown as arrow heads in yellow, lying in the vicinity of the topological defects. Although all MTMT events were localized close to the topological defects (within $\sim$ 1-2 cell lengths), not every topological defect was flanked by an MTMT spot. In general, topological defects emerged throughout the colony, mediated by the spatial distribution of the nematic microdomains, however not every defect supports the onset of an MTMT event in its vicinity. MTMT events occur close to the defects only when defects lie within the P-zone, as shown in Figure \ref{fig:FIG2} (panels A and B). The time at which MTMT occurs, $t_{c}$, depends strongly on the biophysical activity, which in turn, depends on the growth temperature (Figure \ref{fig:FIG1}G). Since the cell aspect ratio of the two strains show similar trends at MTMT (Figure \ref{fig:FIG1}F), we conclude that the transition time for the colony is primarily a function of the growth temperature (and biophysical activity). 

Figure \ref{fig:FIG2}B presents a time sequence of the 2D to 3D extrusion at the instance of cell division, close to the colony center. It should be noted that cell divisions take place also off-center, however the probability of the MTMT event decreases as one move radially outward, since the critical stress criterion is fulfilled only within the finite subsection of the colony called the P-zone. Once the colony transitions to the second layer, both the first and second layers continue expanding, however with different areal speeds. The second later expands faster than the first later; and upon extrusion of the third layer, the third layer expands faster relative to each of the first and second layers (Figure \ref{fig:FIG2}C). As more layers continue adding up (Figures \ref{fig:FIG2}D and E), the 2D areal speed diminishes progressively, while the expansion in the vertical plane picks up. Ultimately, the net horizontal expansion ceases, and the bacterial colony continues expanding purely in the vertical plane. 




\begin{figure}[htbp]
\includegraphics[scale=0.134]{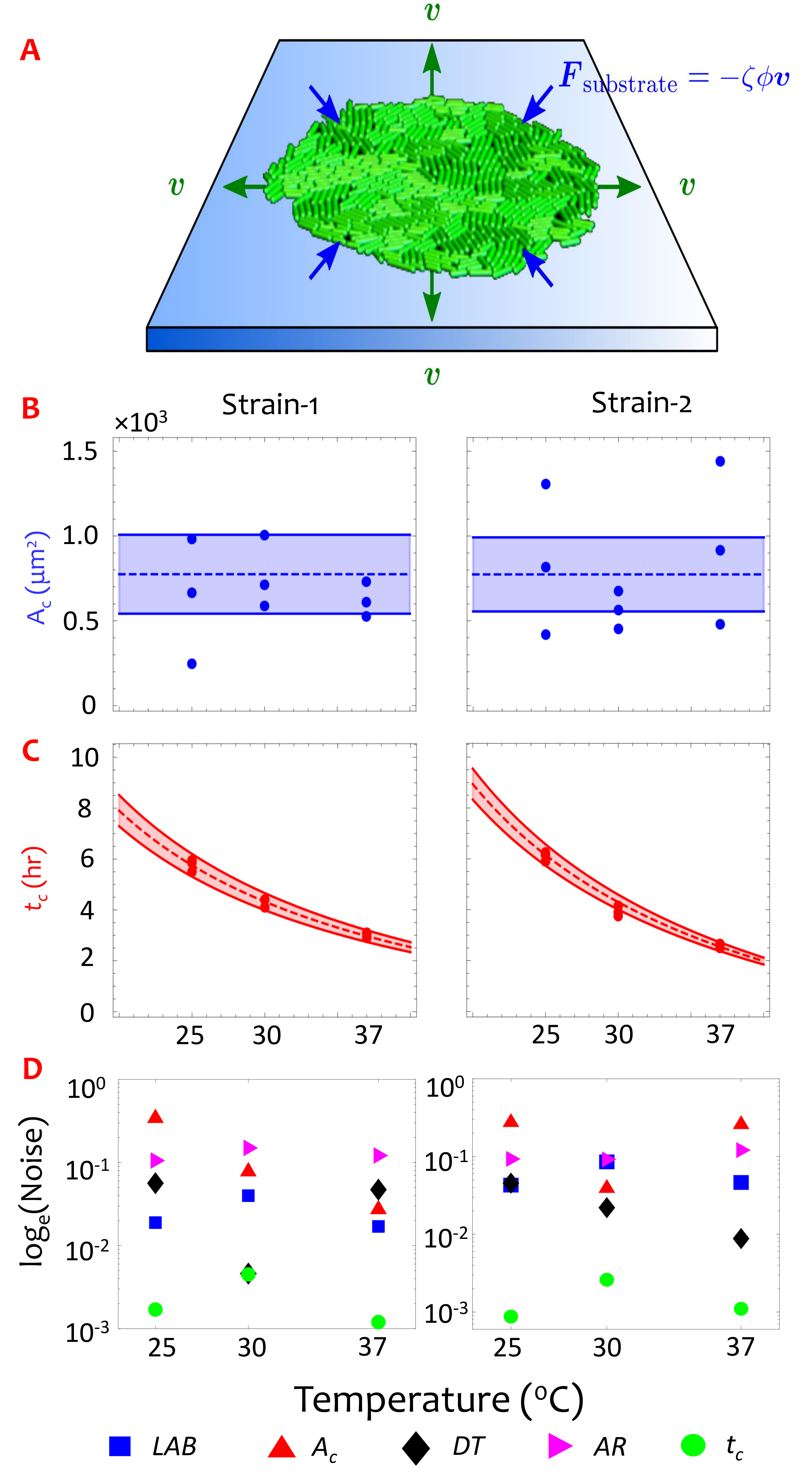} 
\caption{\label{fig:FIG3} \textbf{Trade-offs in phenotypic noise lead to a deterministic transition time to MTMT.}
\textbf{(A)} Schematic representation of the expanding colony underlying our continuum model. The colony grows isotropically under the active forces originating from cell division. Its expansion, however, is hindered by the drag force $\bm{F}_{\rm substrate}=-\zeta\phi\bm{v}$, with $\zeta$ a constant, $\phi$ the cell packing fraction and $\bm{v}$ the local velocity, resulting from the cell-substrate interactions. As a consequence of this obstruction, the cell density in the bulk increases with time, eventually leading to the MTMT.
\textbf{(B)} Critical area at MTMT, $A_c$, is independent of the temperature, captured both in our theory and experiments, whereas the critical time to reach MTMT, $t_c$, is temperature-dependent \textbf{(C)}. The blue and the red points indicate respectively the experimental $A_c$ and $t_c$ data for the two strains. Each data point corresponds to a distinct biological replicate, representing multiple technical replicates. While the variance in $A_c$ is large (distributed randomly within a deterministic range) around the MTMT event, the critical time, $t_c$, exhibits systematic trends with temperature: an overall low variability, which minimizes at the optimum growth temperature (37\textdegree{}C), and a deterministic drop in the $t_c$ as temperature increases to optimal value. In all four plots, dashed lines and shaded regions indicate the predicted mean and standard deviation respectively, see Eqs. \eqref{eq:ac_mean_var} and \eqref{eq:tc_mean_var}. \textbf{(D)} Phenotypic noise, quantified as the normalized variance, $F=\var(\cdots)/\langle \cdots \rangle^{2}$, plotted across different temperatures and strains for the measured traits: LAB (blue squares), critical area at MTMT (red triangles), cell length doubling time (black diamonds), aspect ratio (magenta triangles), and critical time to MTMT (green circles). Despite the high phenotypic noise at individual (LAB, \AR{}, and $DT$) and population scales ($A_c$), the critical time to MTMT ($t_c$) is deterministic with the corresponding noise lying orders of magnitude lower, across all temperatures and strains.
}
\end{figure}

\subsection{\label{sec:ActivityHydTransModel}Trade-offs in phenotypic noise self-regulate critical time to mono- to multilayer transitions}

In Sections \ref{sec:ActivityProxy}-\ref{sec:ActivityVerti}, we elucidate that multiple parameters: \AR{}, LAB, $A$, $N$, along with the growth rates at individual, population- and colony-scales ($\D \ell/\D t$, $\D N/\D t$, and $\D A/\D t$ respectively), specifically at the onset of MTMT, are necessary to  describe the biophysical traits of an expanding colony, ultimately resulting in MTMT. The measured traits elicit log-normal distributions, owing to the intrinsic intra-strain variability, suggesting underlying cross-talks across the biophysical traits.

In order to shed light on the emerging trade-offs in cellular noise during the MTMT, we use a simple continuum model of an isotropically invading colony, whose expansion is solely hindered by a drag force originating from the cell-substrate interactions (see Figure \ref{fig:FIG3}A and Appendix \ref{sec:continuum_model}). In Ref. \cite{You2019}, the authors had demonstrated that extrusion occurs at the scale of individual cells, when these as subject to longitudinal forces larger than
\begin{equation}\label{eq:critical_force}
f_{\rm c} \approx d_{0} k_{\rm a}\ell \;,
\end{equation}
where $d_{0}$ is the cell diameter and $k_{\rm a}\ell$, with $k_{\rm a}$ a constant and $\ell$ the cell length, is the total stiffness of the adhesion complex anchoring a cell to the substrate. As a consequence, newly divided cells are more likely to be extruded once the pressure in their surrounding exceeds the threshold $P_{\rm c}=A_{\rm cap}f_{\rm c}$, where $A_{\rm cap}=\pi d_{0}^{2}/2$ is the area of cells approximately spherical cap. In turn, the pressure field $P=P(r,t)$, with $r$ the distance from the center and $t$ time, varies across the colony proportionally to the cell local packing fraction $\phi=\phi(r,t)$, i.e. $P=P_{0}(\phi-1)$, where $P_{0}$ is a constant independent on the cell growth rate \cite{You2018}. In particular, at the colony's center, where the mono-to-multilayer transition is more likely to occur
\begin{equation}\label{eq:phi_0}
\phi(0,t) \approx \frac{A(t)}{A_{0}}\;,
\end{equation}
where $A(t)=A_{0}\exp k_{\rm d}t$, with $k_{\rm d}=\tau_{\rm d}^{-1}\log 2$ and $\tau_{\rm d}$ the cell doubling time, is the colony's area and the approximation holds at long times (see Supplementary section \ref{sec:continuum_model}  for details). Thus, taking $P(0,t)=P_{\rm c}$ and solving for the area gives the following approximate expressions for the critical area and time
\begin{subequations}\label{eq:critical_area_and_time}
\begin{gather}
A_{\rm c} = A_{0}\left(\ell/\ell_{\rm a}+1\right)\;,\\[5pt]
t_{\rm c} = k_{\rm d}^{-1} \log \left(\ell/\ell_{\rm a}+1\right)\;,
\end{gather}		
\end{subequations}	
where $\ell_{\rm a}=P_{0}A_{\rm cap}/(k_{\rm a}d_{0})$ is a constant length scale, expressing the typical extension of the adhesion molecules when stretched by the force $P_{0}A_{\rm cap}$.

Eqs. \eqref{eq:critical_area_and_time} imply that the statistics of the critical area and the critical time are entirely determined by the probability distribution of the cellular length $\ell$, which, in turn, is log-normal as shown in Figure \ref{fig:FIG1}D. Thus, normalizing $\ell$ by $\ell_{\rm a}$ and taking
\begin{equation}\label{eq:pdf}
P(\ell/\ell_{\rm a}) 
= \frac{1}{\sqrt{2\pi\sigma^{2}}\,\ell/\ell_{\rm a}}\,\exp{\left[-\frac{(\log \ell/\ell_{\rm a}-\mu)^{2}}{2\sigma^{2}}\right]}\;,
\end{equation}
with $\mu = \langle \log\ell/\ell_{\rm a} \rangle$ and $\sigma^2= \text{var}(\log\ell/\ell_{\rm a})$, yields the following approximation for the mean value and variance of the critical area, i.e.
\begin{subequations}\label{eq:ac_mean_var}
\begin{equation}
\langle A_{\rm c} \rangle = A_{0}\bigl(\left\langle \ell/\ell_{\rm a}\right\rangle + 1 \bigr)
\end{equation}
\begin{equation}
\text{var}\ A_{\rm c} = A_{0}^{2}\text{var}(\ell/\ell_{\rm a}),
\end{equation}
\end{subequations}
and time, i.e.
\begin{subequations}\label{eq:tc_mean_var}
\begin{gather}
\langle t_{\rm c} \rangle = k_{\rm d}^{-1}\,\log\bigl(\langle \ell/\ell_{\rm a}\rangle+1\bigr)\;,\\
\text{var}\ t_{\rm c} = \left(\frac{k_{\rm d}^{-1}}{\langle \ell/\ell_{\rm a}\rangle+1}\right)^{2}\text{var}(\ell/\ell_{\rm a}),
\end{gather}
\end{subequations}
where we have expanded $t_{\rm c}$ at the linear order about $\langle \ell/\ell_{\rm a} \rangle$ to obtain Eqs. \eqref{eq:tc_mean_var}. Then, using Eq. \eqref{eq:pdf} allows us to explicitly compute the mean and variance of $\ell/\ell_{\rm a}$ in the form
\begin{subequations}
\begin{gather}
\langle \ell/\ell_{\rm a} \rangle = e^{\mu + \frac{1}{2}\,\sigma^{2}}\;,\\
\text{var}\ \ell/\ell_{\rm a} = e^{2\mu + \sigma^{2}}\left(e^{\sigma^{2}}-1\right)\;.
\end{gather}
\end{subequations}	
Finally, since the cellular length grows exponentially in time until reaching twice the length $\lb$ of the cells at birth, i.e.
\begin{equation}\label{eq:ell}
\ell(t) = 2^{t/\tau_{\rm d}}\lb\;,
\end{equation}
the quantities $\mu$ and $\sigma^{2}$ are, in principle, determined by the statistics of $\lb$ and $\tau_{\rm d}$. Yet one can readily show that, for sufficiently long times, the distribution of the doubling time is unimportant, whereas the mean and variance of the cell solely depend on the statistics of $\lb$. Specifically, normalizing  both both sides of Eq. \eqref{eq:ell} by $\ell_{\rm a}$ and taking the logarithm gives
\begin{equation}\label{eq:log_of_ell}
\log \ell/\ell_{\rm a} = \log \lb/\ell_{\rm a} + (t/\tau_{\rm d})\,\log 2\;.
\end{equation} 
Now, for $t\gg\tau_{\rm d}$, the colony features a large number of cells, whose age is sufficiently diversified in order for their length to span the entire range $\lb
\le \ell \le 2\lb$ and the time $t$ in Eqs. \eqref{eq:ell} and \eqref{eq:log_of_ell} can be treated as a uniformly distributed random variable in the range $0\le t \le \tau_{\rm d}$. As a consequence, the ratio $t/\tau_{\rm d}$ is a uniformly distributed random variable in the unit interval, from which $\langle t/\tau_{\rm d} \rangle = 1/2$ and $\text{var}(t/\tau_{\rm d})=1/12$. Hence
\begin{subequations}\label{eq:mu_sigma}
\begin{gather}
\mu = \langle \log \lb/\ell_{\rm a} \rangle + \frac{1}{2}\,\log 2\;,\\
\sigma^{2} = \var \lb/\ell_{\rm a} + \frac{1}{12}\,(\log 2)^{2}\;.
\end{gather}	
\end{subequations}
In summary, our simple theoretical analysis shows that the statistics of the critical area $A_{\rm c}$ and time $t_{\rm c}$ is ultimately determined by the probability distribution of the length at birth, which, in turn, is log-normal, consistently with previous experimental and theoretical work (see e.g. Refs. \cite{Campos:2014,Amir:2014}). Since temperature mainly affects the cell growth rate, while leaving the statistics of the length at birth essentially unaltered (Figure \ref{fig:FIG1} D and F), the latter consideration suggests that, regardless of the specific bacterial strain, both the average critical area $\langle A_{\rm c}\rangle$ and its standard deviation $\sqrt{\var A_{\rm c}}$ are independent on temperature, whereas $\langle t_{\rm c} \rangle$ and $\sqrt{\var t_{\rm c}}$ decreases like $1/T$, since $k_{\rm d} \sim T$ (Figure \ref{fig:FIG1}D).
In order to assess the significance of these predictions, we compare in Figure \ref{fig:FIG3}B the predicted and measured mean critical area and time for each of the strains. The blue (red) shaded regions are delimited by the corresponding standard deviation, obtain from the square-root of Eqs. (\ref{eq:ac_mean_var}b) and (\ref{eq:tc_mean_var}b). Consistently with our predictions, the mean critical area $\langle A_{\rm c} \rangle$ is independent on temperature, whereas the mean critical time $\langle t_{\rm c} \rangle$ exhibits a prominent $1/T$ dependence.

Figure \ref{fig:FIG3}C presents the intrinsic variability across different biophysical parameters measured in our experiments. We quantify the phenotypic noise as the normalized variance, i.e. $\var(\cdots)/\langle \cdots \rangle^{2}$, of five different quantities, namely LAB, \AR{}, $A_{\rm c}$, $t_{\rm c}$ and $\tau_{\rm d}$ across various temperatures and strains. Notably, $F$ spans nearly three orders of magnitude. We find that, across the strains and growth temperatures, the noise is maximum for critical area at MTMT (red triangles, $F \sim 1$), whereas the critical time to MTMT (green circles) consistently show least noise, with $F\sim10^{-3}$. For other parameters, the phenotypic noise remain bounded by the noise in the critical area and critical time, thus ranging within $10^{-3}<F<1$. It is worth noting that the noise in the length-at-birth (LAB) and the cell aspect ratio (\AR{}) remain nearly constant across all temperatures (for both strains, the noise in LAB and \AR{} fall around $10^{-1.5}$ and $10^{-1}$ respectively). Finally, the noise in $\tau_{\rm d}$, the cell length doubling time, shows a strain-specific variation: for Strain-1, the noise spans $10^{-2.5}<F<10^{-1}$, whereas for Strain-2, the noise decreases as optimal growth temperature is reached ($10^{-2.1}<F<10^{-1.5}$). The deterministic nature of the MTMT critical time--arising from an interplay of \textit{noisy} biophysical traits--suggests a trade-off, specifically in phenotypic traits across growth temperatures that lead to self-regulation of the timing of the MTMT events. Based on our experimental data, we propose that a trade-off can emerge between the variance in single-cell length growth rate ($k_{sc}$) and the single-cell length distribution ($\ell/\ell_\text{a}$), since $\ell/\ell_\text{a}$ $\sim f$($k_{sc}$). The intrinsic dependence of individual cell lengths on the corresponding growth rates reduces the variance in $t_{c}$ as the temperature-dependent growth rate increases, ultimately resulting in a trade-off that self-regulates the MTMT time.

\begin{figure*}
\includegraphics[scale=0.21]{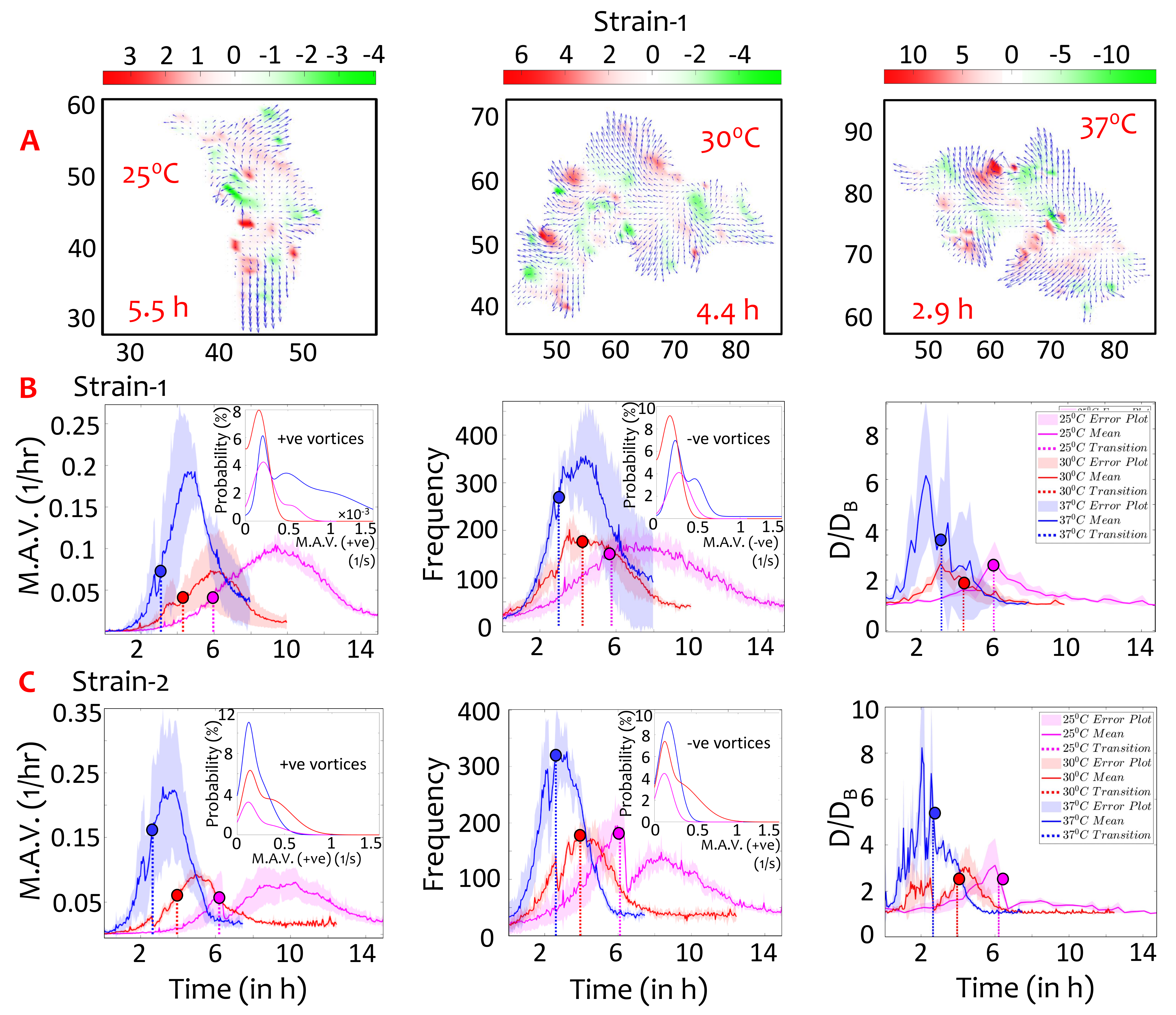} 
\caption{\label{fig:FIG4} \textbf{Activity-dependent emergent hydrodynamics of the confluent colonies is synchronized with the MTMT.}
\textbf{(A)} Growth-induced active stresses trigger velocity and vorticity fields within expanding sessile colonies, reflecting active hydrodynamic patterns in the colony. Vorticity fields just prior to MTMT in Strain-1, at 25\textdegree{}C, 30\textdegree{}C and 37\textdegree{}C, are shown (left to right). The corresponding divergence fields are plotted in Figure \ref{fig:suppFig6}. A movie of the evolving velocity fields is given in Section \ref{sec:Movies} \textbf{Movie S1}. The vorticity strength increases with the biological activity, $i.e.$, with increasing growth temperature, the hydrodynamic activity of the colony becomes stronger. \textbf{(B)} Mean absolute vortices (first column), and the number of unique vortices (second column) for Strain-1, show clear dependence on growth temperature. Corresponding distributions for Strain-2 are plotted in panel \textbf{(C)}. The effective diffusion ratio (third column), plotted against the colony age for the three different temperatures, increase with the biophysical activity. In each plot, the shaded patch denotes the standard deviation error in the observation while the mean is depicted by a solid line. The dashed lines denote the corresponding MTMT event. For the estimation of the enhanced diffusion coefficient, particle radius of 100 nm and medium viscosity of 50 Pas is used (details in the Supplementary section). The insets in panel \textbf{(B)} and \textbf{(C)} depicts the phase-plot of positive and negative vortex distribution. The plots for all the replicates are detailed in Figure \ref{fig:suppFig7}.
}
\end{figure*}

\subsection{\label{sec:ActivityVertiTrans} An expanding confluent colony is a synchronized multifield topological system}

The mono- to multilayer structural transitions trigger active hydrodynamic flows within the bacterial colonies. Figures \ref{fig:FIG4}A-C describe the behaviour of the emergent flows, in relation to the  temperature-mediated biophysical activity. The flow-field, vorticity and relevant distributions are obtained from the time lapse data by adapting a particle image velocimetry (PIV) algorithm to the time lapse colony data technique (refer to the Methods, and the Supplementary sections, and Figures \ref{fig:suppFig5} and \ref{fig:suppFig6}). Post-PIV one obtains a field of flow quantities like velocity, vorticity, strain rate which are computed to an accuracy equal to the floating point accuracy of the system. Since such small differences are indistinguishable in experiments, we set a minimum detection threshold of 10\textsuperscript{-5}rad-s\textsuperscript{-1} for the vorticity, that emerges from scaling arguments. Since the characteristic speed of the colony growth falls in the order of 10\textsuperscript{-9}m-s\textsuperscript{-1}, and the relevant length scale is in the order of 1 $\mu$m, a vorticity scale of 10\textsuperscript{-3}rad-s\textsuperscript{-1} is arrived at. Conservatively, we set our threshold two orders lower to track physically distinguishable vortices. With this consideration, we extract the number of unique vortices (all vortices of the same value when the above vorticity cut-off is applied) from the PIV results and plot it against time in Figure \ref{fig:FIG4}B,C (middle panels). Unique vortices show the perturbations within the system by enumerating the amount of microscale vortices that form due to the colony growth. The results are consistent with the mean absolute vortices, however, the predictive capability of colony transition seems more accurate with unique vortices representation and remains consistent across strains and temperature of the colony growth. The peak of each of the plots lie close to the transition time (dashed line) of the corresponding colour.

Figure \ref{fig:FIG4}A quantifies the vorticity fields of Strain-1 colony close to the MTMT at different growth temperatures. The patchiness in the vorticity field denotes the localization of stress-mediated activity, due either to the local bacterial growth, or the reorganization of nematic microdomains due to cell divisions elsewhere. The strength of the vortex is highest for 37\textdegree{}C, and goes down progressively for colonies grown at lower temperatures (30\textdegree{}C and 25\textdegree{}C). Similarly, the divergence shows a maximum at 37\textdegree{}C (Figure \ref{fig:suppFig6}. The mean absolute vorticity, MAV, underpins the hydrodynamic activity within the expanding colony as shown in Figure \ref{fig:FIG4}B and C (left panels), respectively for Strain-1 and Strain-2. Physically MAV, equivalent to the normalized enstrophy \cite{Lin2021Energetics}, signifies the growth-induced kinetic energy input to the expanding colony. Overall, positive and negative vortices are equally distributed across the colonies, showing no specific handedness over the course of colony development (Figure \ref{fig:suppFig6}). Concomitantly, the net mean divergence remains positive (Figure \ref{fig:suppFig6}), indicating that the bacterial layers without direct contact to the underlying nutrient-rich substrate, still have the possibility to access nutrients from the surrounding micro-environment (following mass conservation). This is discussed further in Section \ref{sec:ActivityHydTrans}. Conspicuously, the MAV peak follows the MTMT event (shown by the dotted vertical lines in Figure \ref{fig:FIG4}B,C), reminiscent of the peak of the areal speeds which follows the MTMT (Figure \ref{fig:FIG1}G). Thus, temporally, the structural and hydrodynamic transitions remain synchronized relative to the timing of the MTMT event. We verified this by reducing the growth temperature, which delayed appearance of the MAV peak, $i.e.$, the time interval between the MTMT and peak MAV increases as the biophysical activity decreased. An interesting facet can be captured upon comparing the MAV at 25\textdegree{}C and 30\textdegree{}C: Since these do not represent optimum growth temperatures, the peaks are similarly valued, however the rates of change of MAV are different, attributable to the biophysical activity. 

The log-normal distributions of the cell aspect ratio and the vortex strength (see Figure \ref{fig:suppFig7}) allow us to connect the topology in structure and active hydrodynamics, and thus reveal a global topological synchrony at time of MTMT. Albeit locally, the cell aspect ratio and vortex strength may remain uncorrelated, our results elucidate links between the topological structure of the active nematic phase and the geometry of the emergent flow fields. An expanding confluent bacterial colony thus represents a dynamical multifield topological system \cite{Giomi2017} where the individual and colony geometry, and the active hydrodynamic flows--both in the colony and its micro-environment--remain temporally and spatially correlated due to the underlying topological attributes of this dynamical system.

\begin{figure*}
\includegraphics[scale=0.125]{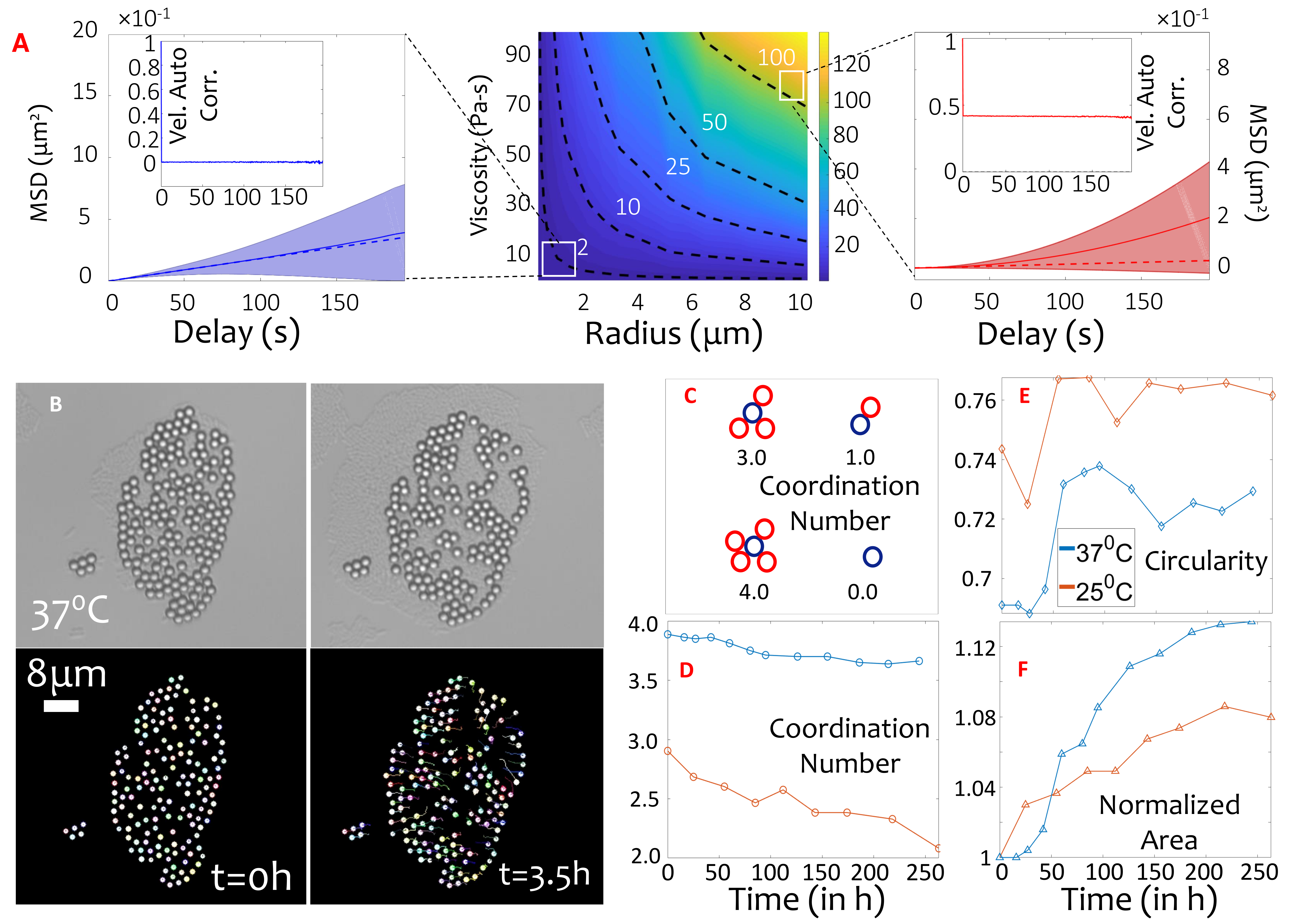} 
\caption{\label{fig:Fig5} \textbf{Confluent colonies actively enhance transport in their micro-environment.}
\textbf{(A)} Simulated transport of passive micro-cargo, based on the experimental results (Figure \ref{fig:FIG4}), described using a phase-plot of the mean squared displacement (MSD) analysis. The gain in the diffusive transport is computed by varying the temperature-regulated biophysical activity experienced by passive particles of different dimensions within a viscous medium. The contour of the gain ratio is delineated using the dotted-black lines with corresponding values in white. At higher viscosity and higher particle radius, the transport is more effective than the Brownian diffusion. The result of two such MSD analysis and the corresponding velocity correlation at two different combination of radius-viscosity situation are provided in the left and right zoomed out sections.
\textbf{(B)} Diffusion coefficient for 25\textdegree{}C and 37\textdegree{}C comes out to $1.3\times10^{-17}$ and $5.1\times10^{-17} m^2/s$. To compare with pure Brownian diffusion at these respective temperatures, the diffusion coefficient is of the order of $10^{-16} m^2/s$. As we have predicted, the diffusion coefficient due to bacterial motion will be negligible for viscosity less than 1 Pas (here agar has a viscosity of around ~0.05-0.1 Pas). A key outcome of the experiments is related to the particle aggregation. If we consider particle aggregation (particle interaction via attractive Van der Waals forces, then the Brownian diffusion becomes much less than $10^{-17} m^2/s$ (Brownian diffusion coefficient with complex particle, of simple shapes, interaction is still an analytically challenging problem). Our control experiments shows that pure Brownian motion is unable to break the particle aggregations, which are easily dislodged by the colony growth. Thus, in consideration with the bacterial flow and particle aggregation, our analytical model under-predicts the strength of the colony growth-induced transport thereby stating a regime of even stronger transport efficiency under complex environment. Under the situation pure Brownian diffusion fails to impress its impact as compared to situations in simple fluids and isolated particles. In biology, most environment have complex fluid structure. \textbf{(C)} Description of coordination number which is the number of particles (in red) attached to the particle of interest (in blue). Two particles are considered attached when their centre-to-centre distance is less than the sum of the two radius of the particles and a tolerance (which is taken as 10\% of the particle radius). \textbf{(D)} Coordination number of the clusters as evolving with time. As it is observed the coordination number decreases as the colony grows, the biophysical activity thus can work towards a de-clustering event which was otherwise not possible by pure Brownian diffusion. \textbf{(E) and (F)} Delineates the circularity and the normalized area of the particle cluster. Due to colony spread, an increase in the circularity (defined as the ratio of the extreme points in the cluster) denotes that overall there is not only a de-clustering phenomena, but the particles are transported equally over all directions. The encompassed ellipse area thus also increase as the de-clustering phenomena evolves highlighting the extent of transport driven by the growing colony of a non-motile bacterial species.
}
\end{figure*}

\subsection{\label{sec:ActivityHydTrans} Confluent colonies actively enhance transport in their micro-environment}
The ability of expanding colonies to successfully transfer momentum to their viscous micro-environment (to trigger meaningful transport) is debated, owing to the long timescales and high dissipation in sessile bacterial populations. It has recently been predicted, computationally, that bacterial carpets comprising active self-propelled entities could trigger enhanced nutrient availability \cite{Mathijssen2018, Lastra2021}. However, experimentally, such micro-environmental transport remains elusive, not least in sessile colonies of non-motile entities. Here, we first use fluid dynamic simulations--with inputs from our experimentally derived hydrodynamic conditions (Section \ref{sec:ActivityVertiTrans})--to assess enhancements in molecular transport due to the activity-dependent hydrodynamics (details in the Methods section, and Figure \ref{fig:suppFig8}). We validate our predictions through micro-cargo experiments (Figures \ref{fig:suppFig9} and \ref{fig:Fig5}), and provide a theoretical framework of the interplay between growth-stresses and viscous properties that trigger enhanced transport. Our results provide biophysical insights into the implications of enhanced transport on sessile bacterial colonies as they transition to multilayer structures.  

Figures ~\ref{fig:FIG4}B and C (panels on the right) present the enhancement of the particle diffusion due to the growth-induced active stresses, for Strain-1 and Strain-2 respectively. Despite the lack of motility, we found that micro-environmental transport properties were altered, evidenced by the $D/D_{B}$ (the ratio of effective and Brownian diffusion coefficients) lying consistently above 1. For the present case, we considered spherical particles with a diameter of 200 nm while the medium viscosity was set to 50 Pas. The present choice of medium viscosity is found in natural settings like mucus or exopolymeric substances (EPS) where the viscosity can be as high as $10^{3}$ Pas \cite{Gloag2020Biofilms}. We harness the biophysical activity \textit{in silico} (here, due to the temperature-induced growth rates), to tune $D/D_{B}$ systematically by up to $\sim$ 4-folds, thus controlling the emergent micro-environmental transport properties. Interestingly, the peak in $D/D_{B}$ occurred at or after the MTMT event, $i.e.$, over a time window when progressively large number of cells lose direct contact with the nutrient-rich substrate (due to extrusion in the third dimension). This indicates an exquisite programmability of the molecular transport, that enables nutrient access to cells within the multilayer structure. 

Figures ~\ref{fig:Fig5}A describes the enhancement of the effective diffusion coefficient with respect to the default Brownian diffusion in the region of growing colony. The transport activity, estimated through the effective normalized diffusion coefficient, is computed as described in Section~\ref{sec:DiffCoeff}. The transport enhancement is consistent with the trend observed in the areal speed and mean absolute vorticity.  The phase diagram reveals that when the medium viscosity is below 1 Pas or when the particle diameter is below 100 nm, the growing colony fails to impact the micro-environmental transport. However, as the environmental viscosity increases--as is the case due to increased release of EPS at the initiation of biofilms--$D/D_{B}$ enhances non-linearly. For a 10 $\mu$m particle, the enhancement is 10 times, when viscosity increases from $\sim$20 to 70 Pas. In section \ref{sec:ExpSetup}, we show that the effect on the particle diameter and viscosity on $D/D_{B}$, and with decrease in temperature, the transport within the colony reduces, thus limiting a colony's ability to take up nutrients (and grow faster), and thus forming a feedback loop. In natural contexts, EPS or any general mucilaginous environment, possess viscosities as high as $10^{3}$ Pas \cite{Gloag2020Biofilms}, thus making such enhanced transport attributes not only detectable (above the default Brownian activity), but also offer effective means to shuttle nutrients, toxins, bacteriophages or other synthetic micro-cargo around colony micro-environments.  

To validate our computational predictions, we allowed colonies to grow in an environment containing low concentration of particles (mean diameter $\approx$ 2$\mu$m), which served as tracers for tracking the enhanced transport properties. As shown in Figures \ref{fig:Fig5}(B-F), we tracked the particles using time lapse imaging (please refer to Supplementary section for details). Remarkably, we observed that the bacterial colony growth has strong influence on micron-sized particles, and despite the low micro-environmental viscosity, ($\sim$0.25 Pas), the bacterial movement could significantly displace the particles (Figure \ref{fig:Fig5}B, see $t= 3.5$ h). We have isolated the particles using image processing and tracked its motion to obtain the effective diffusion coefficient employing the mean squared displacement method. One practical observation that attracted our attention is the ability of the apparently weak bacterial motion to de-cluster passive 2 $\mu$m particles, which otherwise remain strongly aggregated if only Brownian motion persists. By quantifying the particle coordination number over time (Figure \ref{fig:Fig5}, panels C and D), we see a clear drop in the inter-particle connectivity, which depends on the growth temperature. Alongside, the particle-envelope circularity--a relative measure of the transferred momentum--reveals that the circularity increases, however plateaus below 1 (Figure \ref{fig:Fig5}E). A final value $\sim$1 would suggest that the momentum gets isotropically transferred to the passive micro-cargo. Finally, by tracking the normalized area of the particle spread, we see a temperature-dependent increment, capturing clearing the bacteria-induced de-clustering of the colloidal clusters over time (Figure \ref{fig:Fig5}F). In realistic life scenarios, our experimental results could have exciting implications for the active transport of active or passive entities (e.g., phage, nutrients or particulates etc.) which are associated with colonies of sessile bacterial colonies or other confluent cellular systems. 



\section{\label{sec:ActivityOverall}Discussion}

Using a combination of experiments, simulations and continuum modelling, we have quantified the interplay among cellular geometry, topological defects, flow and transport properties in colonies of sessile bacteria at the onset of biofilm formation. To date, how emergence is regulated by the biophysical activity, in relation to the intrinsic noise in cellular phenotypic traits, has remained unexplored. Our time lapse experiments, quantitative image analyses, numerical simulations, and a simple mathematical model, capture the individual-to-colony emergent properties within sets of two distinct {\em E. coli} bacterial strains, growing independently under nutrient replete conditions. We employ growth temperature to systematically tune the biophysical activity, and quantify the phenotypic noise and its impact on the geometry and topology of the emergent structures, flows and transport properties. By mapping the multi-parameter results, across strains and temperatures, we uncover the underlying spatial and temporal interrelations. Growing colonies exhibited strain- and temperature- dependent variability, which we quantify as phenotypic noise (for instance, in single cell and colony growth rates, cell aspect ratio and length-at-birth, or the critical area at MTMT), yet crucially, we discover that the mono- to multilayer structural transitions are deterministic and well-timed, which trigger synchronized transitions in accompanying hydrodynamic traits of the colonies and their micro-environments. 
The active synchrony manifests as one-to-one mapping between the growth rate (biophysical activity) and hydrodynamics, as discussed in Sections \ref{sec:ActivityVertiTrans} and \ref{sec:ActivityHydTrans}. Furthermore, as the trade-off between the noise in the growth rate and the cell aspect ratio (Section \ref{sec:ActivityHydTransModel}) leads to a \textit{deterministic} MTMT time, an overall synchrony across the MTMT and hydrodynamics, and the resulting microscale transport, is concluded.

The geometric and topological interplay ultimately translates into a four-fold (or more) enhancement of the molecular transport in the vicinity of the bacterial colony, despite the cells being non-motile. A simple mathematical model captures the link between the phenotypic noise, and mutual trade-offs therein, that lead to deterministic critical timescales to MTMT transition. The role of activity extends our classical understanding of growth dynamics, specifically in the context of phenotypic noise, and highlights its influence on the emergent colony attributes, both geometric and topological. By harnessing temperature as a tunable cue to regulate the properties of structure and flows associated with sessile bacterial colonies, we establish a generalized mechanistic link between biological activity and emergent properties. Structural and multifield topological attributes discussed in our work can be extended to other dynamic confluent systems, including \cite{Saw2017, Grobas2021MTMTStressors,Turiveaaz2020MTMT,Guillamat2020bioRxiv}. This work offers a rich perspective for future investigations--both in microbial active matter and related cellular systems--by allowing a deeper understanding of growth-mediated transport, with far-reaching ecological, biomedical and environmental ramifications.

\begin{acknowledgments}
This work was supported by the Luxembourg National Research Fund, under the PRIDE Doctoral Training Unit (project MICROH: PRIDE17/11823097), the AFR-Grant (Grant no. 13563560), and the ATTRACT Investigator Grant (Grant no. A17/MS/11572821/MBRACE to AS). This work received generous support from the Human Frontier Science Program  Cross  Disciplinary Fellowship (to  J.D., LT000368/2019-C), and the ERC-CoG grant HexaTissue and by Netherlands Organization for Scientific Research (NWO/OCW) to L.G. We thank Paul Wilmes for the \textit{E. coli} C600-WT strain, and for the lively discussions during the course of the project. Additional thanks to Camille Martin-Gallausiaux and Luise Kleine-Borgmann for microbiological, and Javad Najafi and Christian Wagner for analytical suggestions during the early phases of this work. The authors thank Ghazaleh Eshaghi for her timely assistance with the microparticle assays for the molecular transport experiments.  
\end{acknowledgments}

\section*{Author Contributions} 
AS conceptualized the project, coordinated its operations, and provided overall direction of progress. AT and AS carried out the bacterial growth experiments; JD and AS obtained the microparticle experiments. JD carried out the image processing, PIV analysis, and molecular transport simulations. JD ans AS interpreted and presented the analyzed data. AG provided support with PIV analysis and verified cell geometry data through independent quantification. LG developed the continuum model with specific inputs from JD and AS. JD prepared the experimental plots and figures; LG prepared the computational plots. JD, AG, LG and AS wrote the paper.  




\bibliography{apsJD}


\setlength{\parskip}{20pt}
\newpage

\beginsupplement 
\section*{Supplementary Material}

\setlength{\parskip}{12pt}

\subsection{\label{sec:ExpAndMethods}Methods}

\subsubsection{\label{sec:ExpSetup} Bacterial cultures and bacteria-micro-particle assays} 
 
\textit{Bacterial cultures:} We use two non-motile strains of \textit{E. Coli} bacteria, namely C600-WT and NCM3722 (referred to as the Strain-1 and Strain-2 respectively). As a first step, the cells where streaked on standard agar plates replete with Lysogeny broth (LB). The plated cells were grown for a day, after which isolated cell colonies were identified, and scraped using a microbiological loop. Depending on the downstream experimental requirement, the growth temperature, in each step of cell culturing, was set to one of the temperatures considered in this work (27\textdegree{}C, 30\textdegree{}C, and 37\textdegree{}C). The scraped cells were then transferred to a liquid LB media, and allowed to divide in a shaker for $\sim12$ h. The culture was sub-sampled at regular intervals to track the cell growth over time, using the optical density (OD) measurement technique. After nearly 12 hours of liquid culture growth, the cells were transferred into fresh LB media, at a 1:1000 ratio of cell to fresh media, and grown for $\sim2$ h, before they were introduced on the specially designed substrates (see Figure \ref{fig:FIG1}) for time lapse imaging the colony expansion at 27\textdegree{}C, 30\textdegree{}C, and 37\textdegree{}C.The single cell-to-colony dynamics was observed using time-lapse microscopy on a 2-mm thick layer of agarose gel. The gel is uniformly mixed with LB medium which is a nutrient-rich medium commonly used for growing bacteria under laboratory settings (Figure \ref{fig:FIG1}a). This nutrient-rich layer is sandwiched between two glass slides and a 2-mm thick Gene Frame (spacer) was used to enclose the glass-agarose system. A time lapse phase contrast microscopy imaged the cell dynamics from below. 

\textit{Bacteria-particle assay for quantification of transport properties:} For culturing the bacteria along with the micro-particles, we used 2\(\mu\)m sized polystyrene (initial concentration of 98\% v/v, obtained from Sigma-Aldrich). We first dilute the particles in sterilized de-ionized (DI) water (50\(\mu\)l of the 2\(\mu\)m particle solution in 1 ml of DI water). The mixture is treated vortexed and treated in sonication bath for 2 minutes, and then centrifuged (at 600 rpm) for another 2 minutes. The agglomerated beads are separated and re-suspended in DI water, and the steps are repeated, before introducing the suspension in 1 ml LB medium. The mixture is again sonicated and then centrifuged for 2 minutes each sequentially. The process is repeated three times to ensure the beads are finally suspended in a 1ml LB medium, to reach a \textit{low} final concentration of $\sim10^{3}$ particles/ml. Finally the 1ml LB+beads mixture is mixed with 5ml pure LB medium and sonicated for 2 minutes, so as to hinder the flocculation or particle sedimentation. Hereon, the bacterial strains were cultured in the 6 ml LB+particle medium for each temperature used in the study. For propagation of the bacterial culture at each temperature in the LB+beads medium, we follow the same method as described in \ref{sec:ExpAndMethods} (Bacterial culture). After an interval of 2 hours, we observe the growth of the bacterial strain under the microscope to ascertain its fitness in the medium with beads. Our observation delineates its fitness remains unchanged with the presence of beads. Bio-compatibility was tested by allowing the bacterial cells to grow in this dilute medium over multiple generations. We compared the growth rate, and the geometry of individual cells (microscopy) across all temperatures, and compared them against the control data (cells grown without the micro-particles). No difference could be statistically measured between the two samples sets, thus statistically ruling out an cytotoxic effects due to the dilute particle suspension. After fabrication of the agarose substrate (described earlier, Figure\ref{fig:FIG1}), growth of the colonies and concommitant particle transport are visualized at 60X and 40X magnifications, while maintaining the sample at a particular temperature (see Section on \textit{Time-lapse imaging}). For visualization, we select six colonies for each temperature and noted the X-Y coordinates of their initial positions. At an interval of 30 minutes, the change of the position of the beads is captured (an example is shown in Figure \ref{fig:Fig5}). From the captured images, the micro-particles are identified and their centroids are extracted using image analysis tools. The evolution of the position of the centroid of the beads is extracted using the Mosaic track package in ImageJ. The trajectories of the beads for all the captured colonies at each temperature provides us with the effective diffusion coefficient from the MSD analysis. The evolution of the particle centroids are also used to quantify the coordination number, bounding elliptical area of particle locations and the circularity of the bounding ellipse (see Figure \ref{fig:Fig5} for details).


\subsubsection{\label{sec:TL}Time-lapse imaging}
For experimentation at each particular temperature, we cultured the cells overnight in the LB medium and maintained the culture in a temperature-controlled shaker. For the present study, we maintained the cultures in 25\textdegree{}C, 30\textdegree{}C and 37\textdegree{}C. From the culture, a dilute concentration is extracted and placed on the agarose plate on which a single bacterium was spotted. The subsequent growth of this single bacterium into colonies is imaged while maintaining a temperature corresponding to the growth of the culture within the microscope environment. Such single bacterium acts as the nucleating sites for growth of monoclonal colonies. For each strain and temperature, we have performed three sets of experiments. Statistics over our analysis were measured over all these replicates. For the analysis of MTMT, we have further visualized more colonies in some cases to ascertain its lower variance. We have observed a variability in morphological parameters such as length-at-birth and length-at-transition within colonies even under similar conditions potentially attributable to phenotypic heterogeneity.

The colony growth in two-dimension and subsequent penetration to the third dimension is visualized using time-lapse phase-contrast microscopy. Images were acquired using an 	Hamamatsu ORCA-Flash Camera ($1\ \mu m = 10.55$ pixels) that was coupled to an inverted microscope (Olympus CellSense LS-IXplore) with a 60X oil objective. Overall, this gave a resolution of 0.11 $\mu m$. The microscope was stage was enclosed within a thermally insulated temperature-controlled incubator (Pecon), which could be regulated precisely to set the temperature, and monitored temperature at the sample with a resolution of 0.1\textdegree{}C. Prior to initiation of capturing the images, we identified and recorded multiple locations on the agarose surface where single bacterium was present. The microscope was automated to scan these pre-recorded coordinates and to capture the images of the gradually increasing colonies at every 3 min interval while maintaining the focus across all the colonies captured. The captured and saved images over hours gives us the necessary data to analyse the MTMT, dynamics and transport within the bacterial colonies. We extracted the dimensions (width and length), position (centroid) and the orientation of each bacteria from the phase-contrast images using the combination of open-source packages of Ilastik \cite{berg2019Ilastik} and ImageJ as well as MATLAB (MathWorks), as detailed in Section \ref{sec:ImageAnalysis}. Upon extraction of the cell morphological properties, we are able to generate the orientation maps of the colony  (Figure \ref{fig:FIG1}b).

\subsubsection{\label{sec:ImageAnalysis}Image analysis}
\textit{Image segmentation, cell geometry analysis and cell counting:}The counting process consists of cell segmentation followed by counting the number of individual entities. Cell segmentation is performed using a combination of Ilastik-MATLAB coding that helped to extract the bacterial length and orientation in the colony for each time frame of the bacterial colony growth. The process is continued till the colony encounters MTMT, since after this the image contrast and focus gets limited for subsequent segmentation to be carried out (PIV analysis, below, is still feasible). Initially, pre-processing of the raw images is performed by a combination of background filling and weighted bottom-hat and top-hat filter application \ref{fig:suppFig3}. On the pre-processed image, Ilastik was trained for bacterial segmentation. The training process involves iteration until a reasonably satisfactory extraction is obtained. A labelled image is extracted from the segmentation process and identified in MATLAB. A bacillus-shaped water-shedding technique is performed to separate out joint bacterial cells. Finally, these individual entities (segmented cells) are coloured (Figure \ref{fig:suppFig2}F) or outlined (Figure \ref{fig:suppFig4}) for counting and analysis. The orientation of the individual bacteria leads to an effective director profile of the microdomains emerging with the expanding colony (as shown in Figure \ref{fig:FIG1}B). Details of the orientational analysis for microdomain detection is discussed in Ref.\cite{You2018}. Once the microdomains are tracked, the topological defects are identified as the intersection of three or more micro-domains, and further verified visually for each colony data. Depending on the rotational nature of the change in the microdomain orientations at the intersection (clockwise or counter-clockwise, as for polarization optics of topological defects in ordered materials), the defect sign was assigned as (+)ve or (-)ve. 

\textit{Colony extraction and PIV analysis:} The image processing has been carried out using the MATLAB Image Processing module. Following are the image processing steps we have used to extract the bacterial colony: (in sequence) an adaptive thresholding, image dilating and filling, followed by image labeling. Then the label that coincides with a given centre, which is any point on the first bacterial cell, is extracted and dilated, again followed by image filling and finally an image erosion is applied. A Boolean intersection carried out between the image from the last step and the original (raw) image satisfactorily extracts the outline of the colony at each time frame. The colony outline gives an effective colony area which changes with time and has been tracked for each of the experiments. \par
The PIV analysis is performed on the final extracted colony images (frames which not only precludes the background noise in the flow due to the colony growth and light interference, it also helps to focus on a single colony in cases multiple colony growth were captured in a single frame). For the PIV analysis, initially the Contrast-limited adaptive histogram equalization (CLAHE) filter is applied to each image for better contrast. The FFT-based cross correlation algorithm was found to be optimum with 3-pass interrogation area. Specifically, the desired output from the PIV analysis are the two velocity components and the vorticity field at each time frame which is generated within the bacterial colony due to its spread. The PIV analysis with the processed images are carried out with PIVlab - particle image velocimetry (PIV) toolbox of MATLAB \cite{PIVLabThielicke2014}.

\subsubsection{\label{sec:continuum_model} Hydrodynamic model}

In this Appendix, we provide a derivation of Eq. \eqref{eq:phi_0}. The late temporal dynamics of growing colonies, can be conveniently described upon modelling the colony as a two-dimensional continuum, whose total mass $M=\int {\rm d}^{2}r\,\rho$, with $\rho=\rho(\bm{r},t)$ the density, grows exponentially in time and whose momentum density $\rho\bm{v}$, with $\bm{v}=\bm{v}(\bm{r},t)$ the velocity, evolves under the combined effect of pressure gradients and drag. The corresponding hydrodynamic equations are given by:
\begin{subequations}\label{eq:eom}
\begin{gather}
\partial_{t}\phi + \nabla\cdot(\phi\bm{v}) = k_{\rm d}\phi\;,\\[5pt]
\rho_{\rm cell}\left[\partial_{t}(\phi\bm{v})+\nabla\cdot(\phi\bm{v}\bm{v})\right] = -\nabla P - \zeta\phi\bm{v}\;,	
\end{gather}
\end{subequations}
where $\phi=\rho/\rho_{\rm cell}$, with $\rho_{\rm cell}$ the average density of individual cells, is the local packing fraction, $k_{\rm d}=\tau_{\rm d}^{-1}\log 2$, with $\tau_{\rm d}$ the cell doubling time, $P$ the pressure and $\zeta$ a kinetic drag coefficient. Now, upon neglecting inertial effects, Eq. (\ref{eq:eom}b) can be readily solved to give a Darcy-like expression for the velocity field, namely
\begin{equation}\label{eq:darcy}
\phi\bm{v} = - \frac{1}{\zeta}\,\nabla P\;.	
\end{equation}
Then, using the equation of state introduced in Sec. \ref{sec:ActivityOverall}, i.e. $P=P_{0}(\phi-1)$, taking the divergence of Eq. \eqref{eq:darcy} and replacing the resulting expression in Eq. (\ref{eq:eom}a), yields a single partial differential equation for the packing fraction $\phi$, that is
\begin{equation}\label{eq:phi}
\partial_{t}\phi = D\nabla^{2}\phi + k_{\rm d}\phi\;,
\end{equation}
where $D=P_{0}/\zeta$ is an effective diffusion coefficient. Eq. \eqref{eq:phi} can be readily solved with initial conditions $\phi(\bm{r},0) = A_{0}\delta(\bm{r})$, where $A_{0}$ the initial area of the colony. This gives
\begin{equation}\label{eq:phi_solution}
\phi(r,t) = \frac{A(t)}{4\pi D t}\,e^{-\frac{r^{2}}{4 D t}}\;,	
\end{equation}
where 
\begin{equation}\label{eq:area}
A(t)=\int {\rm d}^{2}r\,\phi(\bm{r},t) = A_{0}\,e^{k_{\rm d}t}\;,
\end{equation}
is area of the colony at time $t$. Now, at the center of the colony one can approximate
\begin{equation}\label{eq:phi_limit}
\phi(0,t) 
= \exp\left(k_{\rm d}t-\log\frac{4\pi D t}{A_{0}}\right)
\xrightarrow{k_{\rm d}t \gg 1}e^{k_{\rm d}t}\;,
\end{equation}
as long as
\begin{equation}
k_{\rm d}<\frac{4\pi D}{A_{0}}\;.
\end{equation}
Finally, dividing and multiplying the right-hand side of Eq. \eqref{eq:phi_limit} by $A_{0}$ and using Eq. \eqref{eq:area} yields Eq. \eqref{eq:phi_0}. Note that areal diffusion $D$ and initial area $A_0$ has order of magnitudes ~10$^{-13}\ m^2/s$ and ~$10^{-11}\ m^2$. This makes the r.h.s. of the ratio in above equation in ~0.1$s^{-1}$ to ~0.01$s^{-1}$. With doubling time being in orders of minutes, the corresponding value of $k_d$ becomes 10$^{-3}s^{-1}$ or lesser, thereby bringing forth an experimental agreement to the above relation.

\subsubsection{\label{sec:DiffCoeff}Data-based transport simulations}

To obtain a data-based effective diffusion coefficient from the colony-mediated hydrodynamics, we use the particle tracing approach wherein a certain number of particles are placed randomly in the flow field induced by the expanding bacterial colony. The position of the particles are then tracked over time. Here the evolution of the particle position is achieved by simplifying the Generalized Langevin Equations (GLE). Force balance using two-dimensional GLE formulations for mass-less particles with a background flow field reads \( \zeta\frac{d\textbf{x}}{dt}=\zeta\textbf{u}+\zeta\sqrt{4D}\textbf{W}(t)\) where \( \textbf{u} \) is the velocity field (induced due to the bacterial growth), \( \textbf{W}(t) \) is a normally distributed random noise satisfying \( <\textbf{W}> = 0 \) (\(<>\) denotes a mean), \( \zeta \) is the fluid friction force coefficient, \( D \) is the diffusion coefficient and \( \textbf{x} \) is the particle position. The velocity is applied to the equation depending on the position the particle occupies at any particular time in the domain. The corresponding probability distribution function \(P(\textbf{x},t)\) as solved from the Fokker-Planck equation is given by \(P(\textbf{x},t)=\frac{1}{\sqrt{4\pi DT}}exp(-\frac{(\textbf{x}-\textbf{u}t)^2}{4Dt})\). The first and second moment of the particle position reads \( M_1(\textbf{x}) = \int_{-\infty}^{\infty} \textbf{x}P(\textbf{x},t) = <\textbf{u}t> \) and \( M_2(\textbf{x}) = \int_{-\infty}^{\infty} x^2P(\textbf{x},t) = <u^2t^2> + 4Dt \) \cite{MSDMedved2020}. The second moment equals the \textit{mean squared displacement} (MSD) of the particle positions. The above relation subtly implies that the MSD is no more a linear function of time. The diffusion coefficient magnitude in the second moment is replaced using the \textit{fluctuation-dissipation theorem} which states the classical Stokes-Einstein formulae as \(D=\frac{k_BT}{6\pi\mu a} \) where \(k_B\), \(T\), \(\mu\) and \(a\) are the Boltzmann constant, absolute temperature, medium viscosity and particle radius, respectively. Using the \(M_2\) relation, we find the evolution of the particle position with time. From these particle trajectories, we employ back the MSD analysis to get the effective diffusion coefficient. The diffusion coefficient is plotted normalized by the pure Brownian diffusion (\(D_B\)) to get a measure of the enhancement in the diffusion due to bacterial colony growth. Besides the normalized diffusion coefficient, we have also plotted the change in the mean particle position (\(M_1\)) and the velocity correlation (in the Supplementary) to get a gauge on the diffusive-like nature of the transport phenomena.  

\begin{table*}
\centering
\caption{\label{tab:parameters} Glossary of symbols used in the study.}
\smallskip
\begin{tabular}{l | c | c}
{\bf Physical parameter} & {\bf Units} & {\bf Parameter symbol} \\
\hline \hline
Temperature & \textdegree{}C &  $T$ \\
Instantaneous colony area &$m^2$ &  $A$\\
Critical area (colony area at MTMT) & $m^2$ &  $A_c$\\
Critical time (time to MTMT) & $s$ &  $t_c$\\
Packing fraction & 1 & $\phi$\\
Number of cells & 1 & $N$\\
Number of defects & 1 & $N_{d}$\\
Defect concentration & $m^{-2}$ & $C_{d}$\\
\hline
\multicolumn{3}{|c|}{Dynamics Parameters} \\
\hline
Pressure & $N/m^2$ &  $P$ \\
Velocity field & $m/s$ &  $v$\\
Critical force & $N$ &  $f_c$\\
Mean averaged vortex & $m^2/s$ &  MAV\\
Kinetic Drag coefficient & $Pas/m^2$ & $\zeta$\\
Effective Diffusion coefficient & $m^2/s$ &  $D$\\
Brownian Diffusion coefficient & $m^2/s$ &  $D_B$\\
Mean Squared Displacement & $m^2$ &  MSD\\
\hline
\multicolumn{3}{|c|}{Phenotypic Parameters} \\
\hline
Aspect Ratio  & 1 &  \AR{}\\
Bacteria length & $m$ & $l$\\
Bacteria width & $m$ & $w$\\
Spherical cap area & $m^2$ &  $A_{cap}$\\
Length scale & $m$ &  $l_a$\\
Initial area & $m^2$ &  $A_0$\\
Cell number growth rate & $s^{-1}$ &  $k_d$ \\
Cell number doubling time & $s$ &  $\tau_d$ \\
Cell length growth rate & $s^{-1}$ &  $k_{sc}$ \\
Cell length doubling time & $s$ &  $DT$ \\
Colony Area doubling time & $s$ &  $t_A$ \\
Mean of log of dimensionless length & 1 &  $\mu$\\
STD of log of dimensionless length & 1 & $\sigma$\\
Mean & unit specific &  $\mu_0$\\
Standard deviation (STD) & unit specific & $\sigma_0$\\
Phenotypic noise & 1 & $F$\\

\hline
\end{tabular}
\end{table*}

\subsection{\label{sec:SuppFigs}Supplementary Tables and Figures}

This section contains the cell counting and doubling time results from various parameters, image processing and cell segmentation steps detailed through graphical presentation, defect identification and quantification, colony extraction for PIV, PIV analysis, phase-plots of vortex strength distribution, tracks of passive particles generated from data-based simulation with which MSD and velocity correlations are estimated and discussion and quantification on particle transport experiments. The Supplementary Table \ref{tab:parameters} gives a glossary of all the symbols used in the article for easy reference. 

\begin{figure*}[htbp]
\includegraphics[scale=0.08]{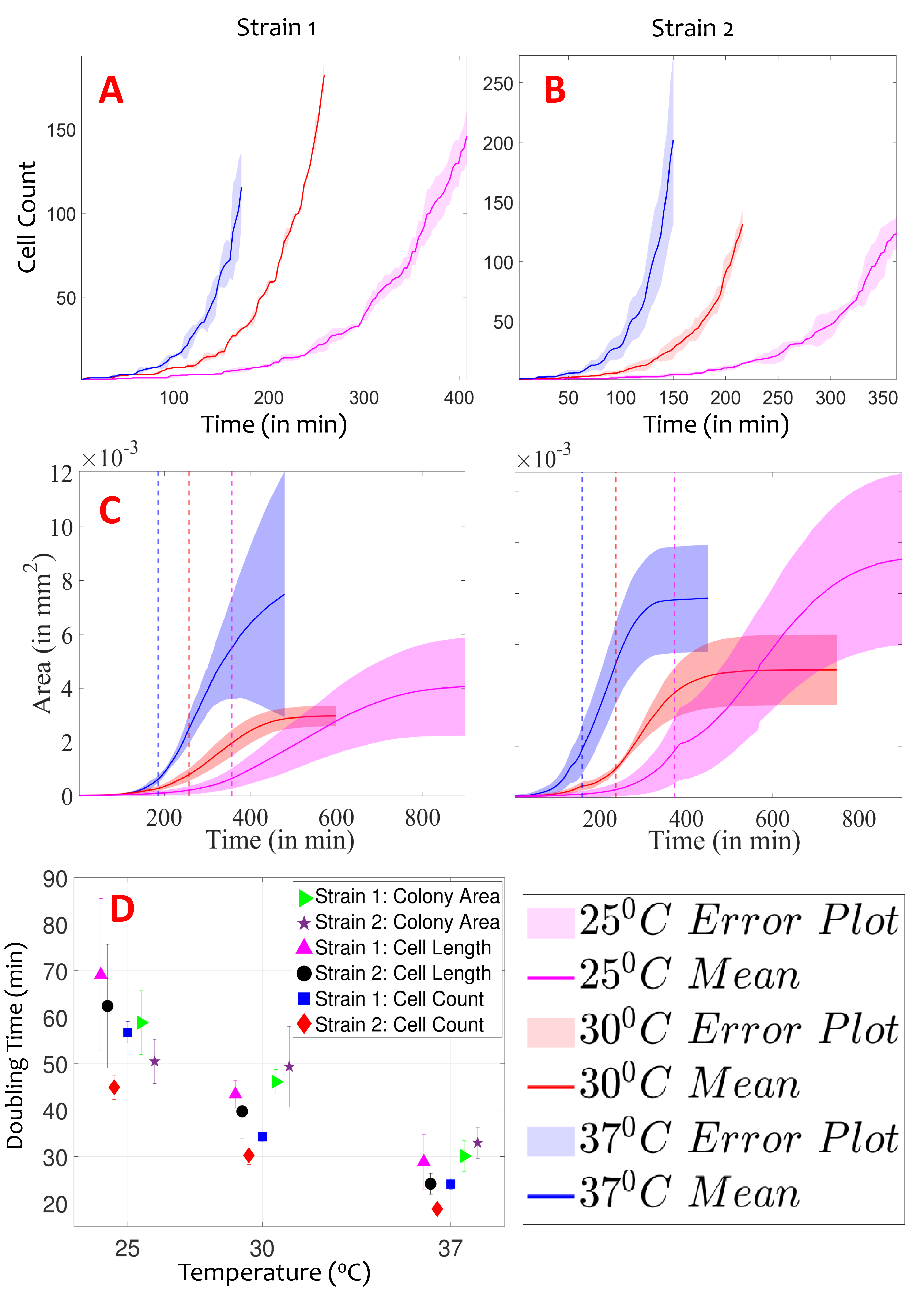}
\caption{\label{fig:suppFig1} \textbf{Growth and doubling parameters.} Depicts the growth for \textbf{(A)} Strain-1 and \textbf{(B)} Strain-2 at different temperatures. The growth efficacy increases with an increase in temperature, i.e. as the temperature approaches the optimum temperature of growth, which is 37$^0$C, the growth of the colony approaches its maximum potential to grow (keeping all other factors constant). \textbf{(C)} The evolution of colony area with time for each strain and temperature. The legend is same as in panel \textbf{(A)}. \textbf{(D)} The three different growth rates as obtained from single cell growth, cell population growth and colony area growth are depicted in this figure. The growth rate is obtained from the exponential stage of the growth regime for cell population and colony area growth. The growth rate values extracted from these three phenomena remains in close proximity with each other signifying how one affects the other. The growth rate, a reflection of bacterial activity, increases with temperature, thus, posing temperature as the proxy for tuning ensuing biophysical, hydrodynamic and transport activity. Supplementary section \ref{sec:Movies} \textbf{Movie S1} A, D and G show the evolution of bacterial colony (representative one replicate) for 37\textdegree{}C, 30\textdegree{}C and 25\textdegree{}C, respectively.}
\end{figure*}

\begin{figure*}[htbp]
\includegraphics[scale=0.15]{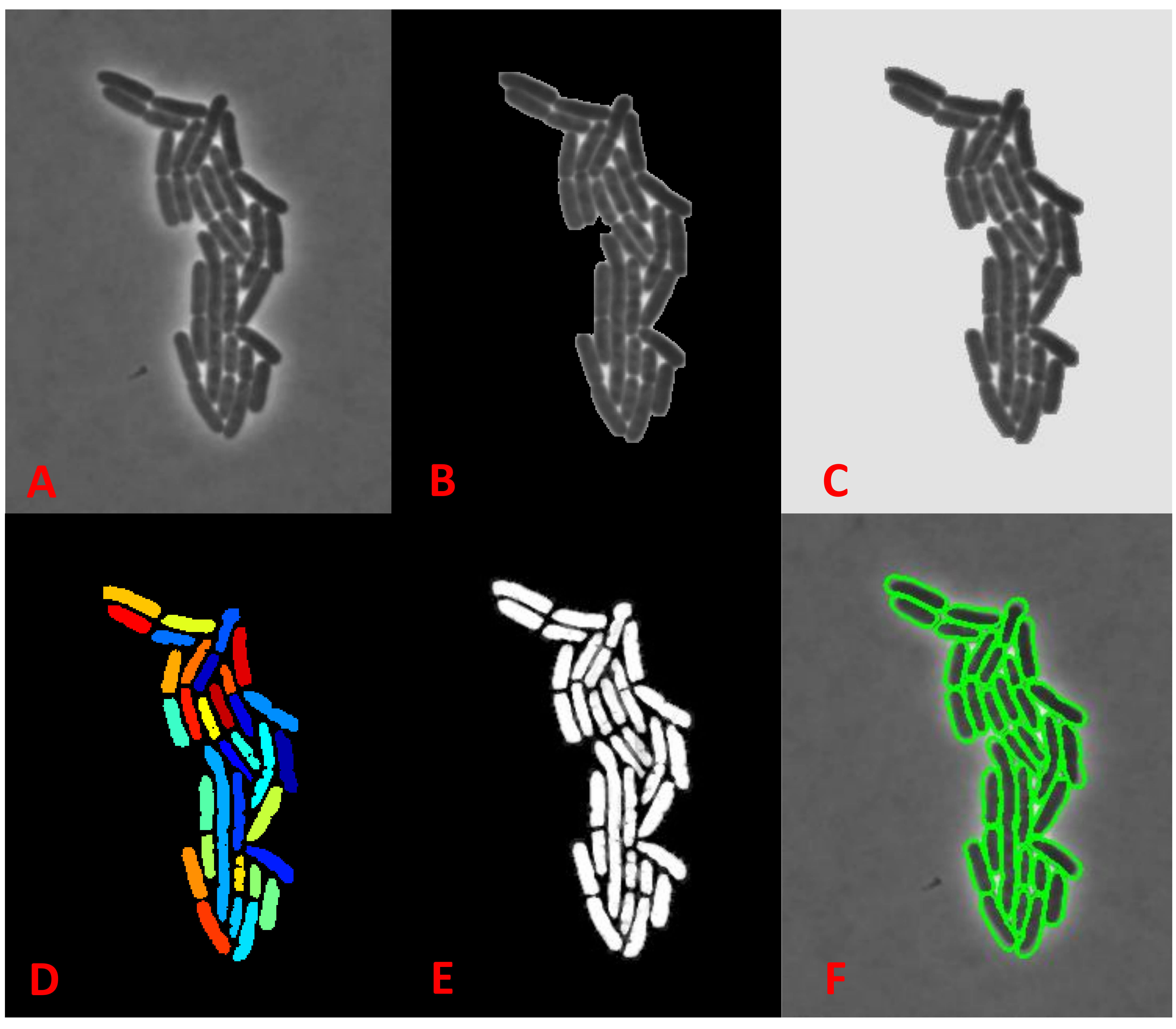}
\caption{\label{fig:suppFig2} \textbf{Cell segmentation and counting.} \textbf{(A - C)} The protocol for counting cells is established using a combination of AI-based Ilastik engine and MATLAB. First, MATLAB was used to pre-process the colony from \textbf{(A)} to \textbf{(C)}, which creates a strong contrast for better training and segmentation programming, details in Figure \ref{fig:suppFig3}. \textbf{(B - D)} Ilastik was used for cell segmentation wherein numerous bacterial cells were used to train the model to properly differentiate between bacterial cells and the background. Once the segmentation labels were obtained, they were identified by MATLAB, followed by the identification, outlining and cell counting steps. The analyzed images were subjected to manual quality check to detect any discrepancy with the automated analysis pipeline. \textbf{(E - F)} Shows the segmented logical images of the cells and the logical image used to outline the cells overlayed on the raw image, respectively. These images gives the cell geometrical details.}
\end{figure*}

\begin{figure*}[htbp]
\includegraphics[scale=0.2]{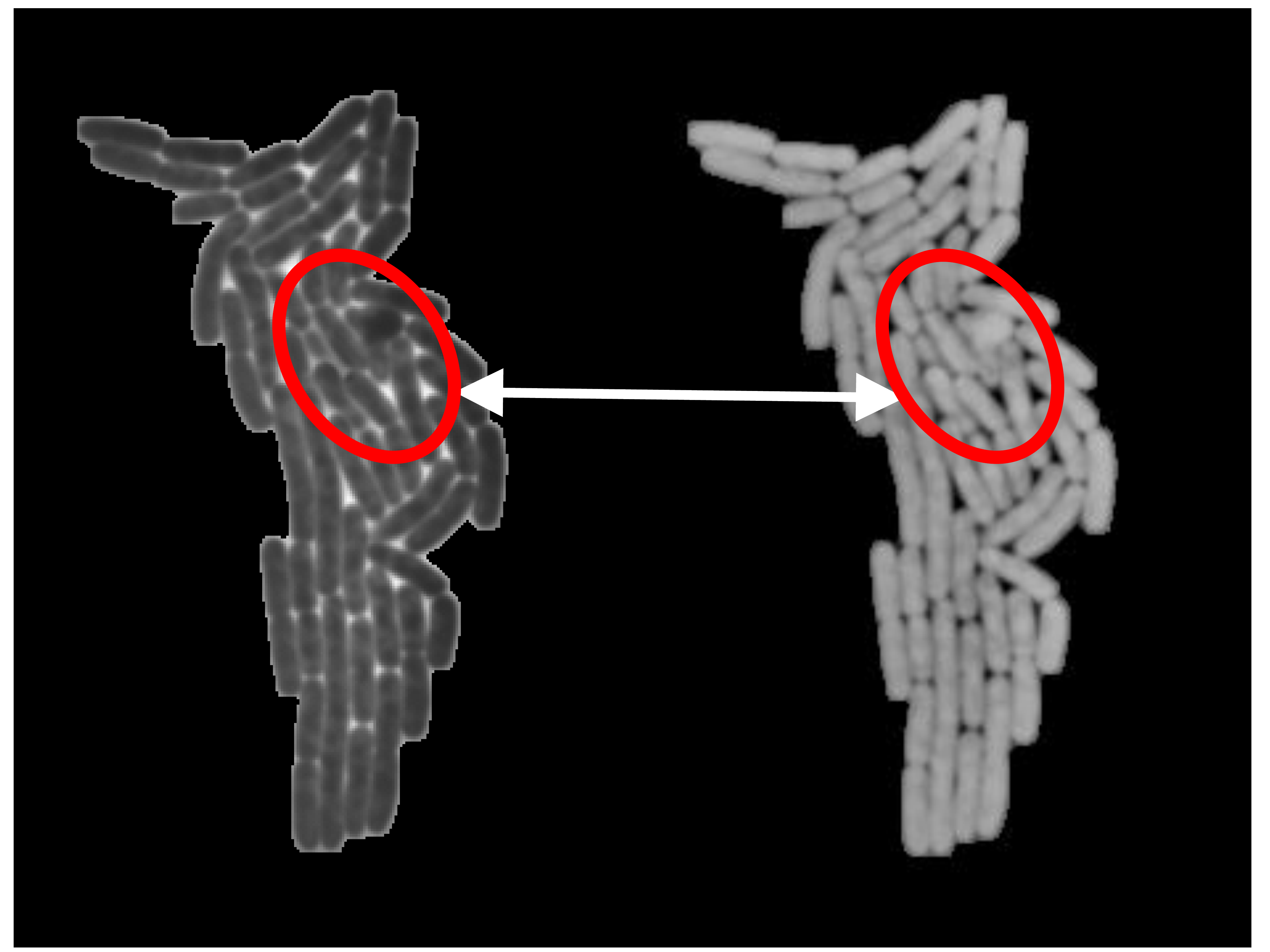}
\caption{\label{fig:suppFig3} \textbf{Tracking the onset of the mono to multilayer transition (MTMT) event.} The formation of a second-layer bacterial structure was identified through a series of image processing steps. In the left panel, we observe the extracted bacterial colony (at time 5.7 hours for a colony growing at 25\textdegree{}C). The right panel shows the colony after application of an image filter (described below) which clearly brings out the pattern of bacterial overlap, $i.e.$, a bacteria growing into a second layer situated at the top of the first layer. For the image filter, first the colony was extracted (Figure \ref{fig:suppFig2}B), it segments the colony from the background (Figure \ref{fig:suppFig5}). The background of Figure \ref{fig:suppFig2}B is filled up with a colour that represents the near-whitish contrast of the bacterial cell and its immediate background (Figure \ref{fig:suppFig2}C). Hereon, a combination of top-hat and bottom-hat filters are applied that eventually brings out the visually contrast of transitioning cells in the figure analyzed. One can clearly distinguish that the bacterial cells start growing on top of each other triggering the phenomenon of monolayer-to-multilayer transition. The present technique is used to extract the location of monolayer-to-multilayer transition along with the critical area and critical time that is used in Figure \ref{fig:FIG3}. All automated image analysis data were subject to manual quality checks, so as to rectify any discrepancy.}
\end{figure*}

\begin{figure*}[htbp]
\includegraphics[scale=0.3]{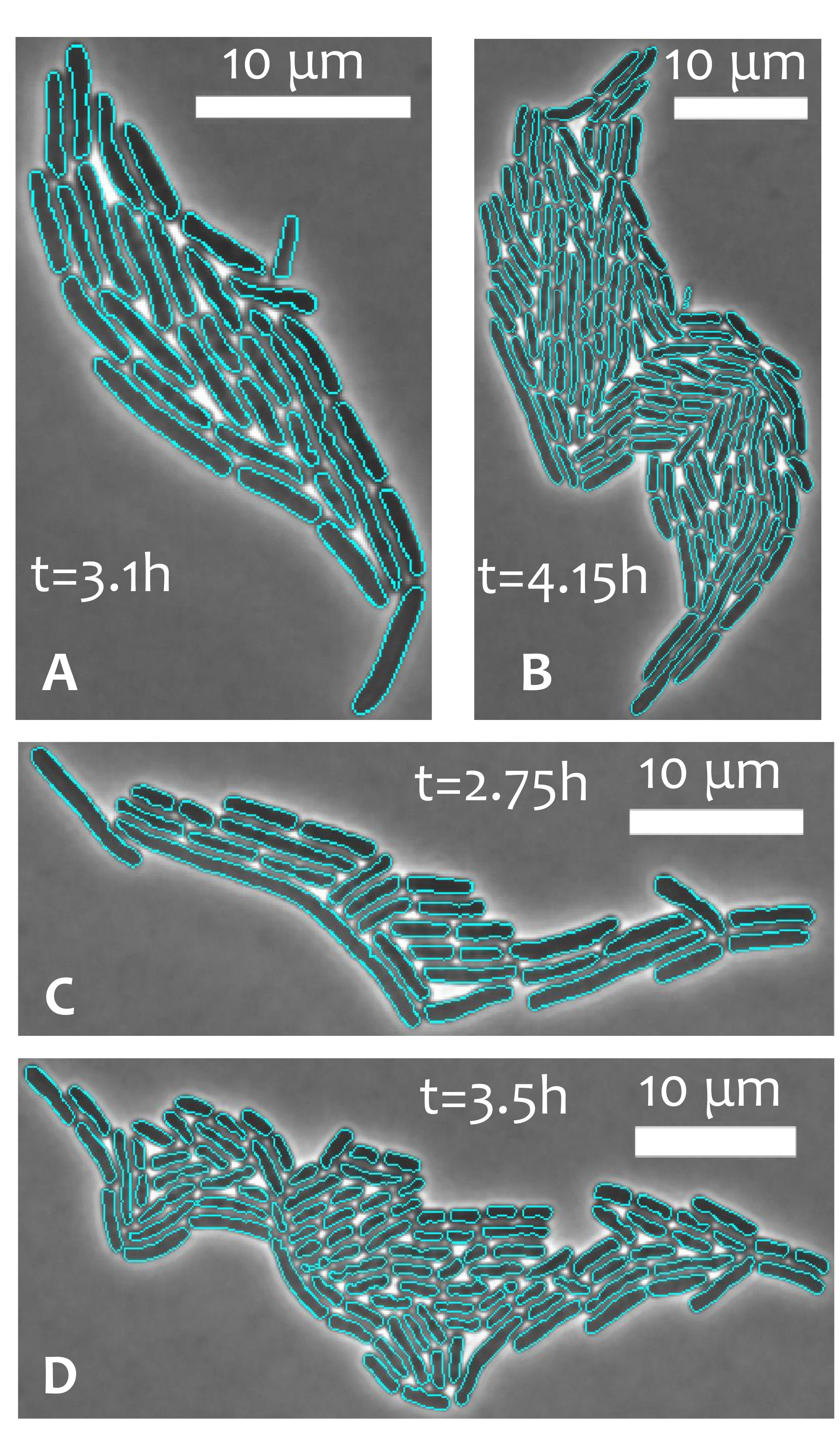}
\caption{\label{fig:suppFig4} \textbf{Quantifying cell aspect ratio.} The comparison of \AR{} at early and late time points in a growing colony of Strain-1, shown in \textbf{(A)} and \textbf{(B)} respectively, for 30\textdegree{}C. One time point is selected that is much before the MTMT event (younger generation) while the other selected time point is just before MTMT occurrence. We segmented the cells and estimated the mean AR at the two time points for each strains. For Strain-1, panels A and B has mean AR of 6.1 and 5.26. For Strain-2, panels C and D have mean AR of 5.9 and 4.1. Thus, the AR decreases as one approaches MTMT.}
\end{figure*}

\begin{figure*}[htbp]
\includegraphics[scale=0.14]{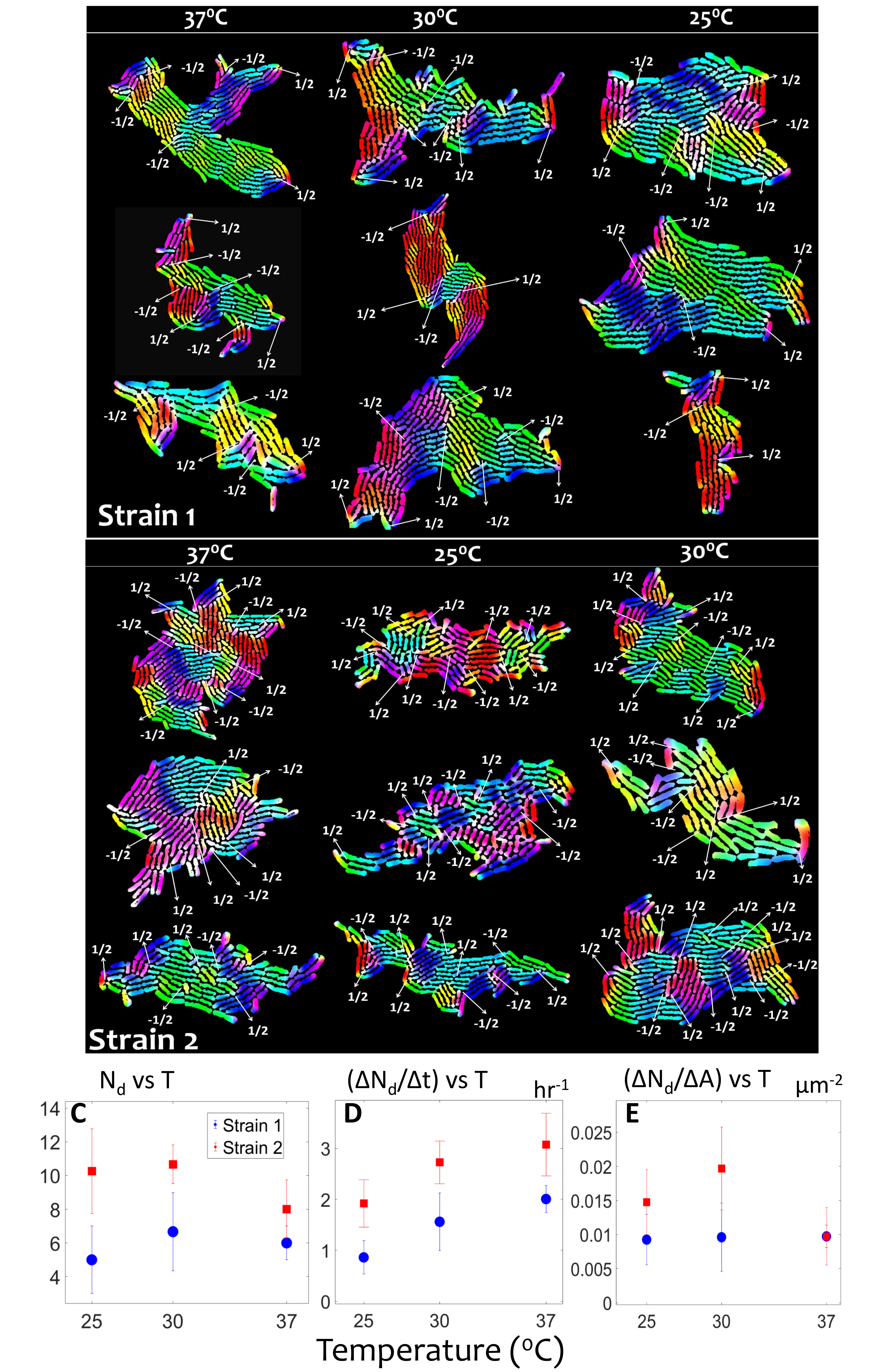}
\caption{\label{fig:suppFig4ab} \textbf{Extracting topological defects at the MTMT event.} \textbf{(A)} and \textbf{(B)} The montage of the defects at MTMT in each of the strains for each replicates and temperature. The details of the image and orientational analysis is described in Section \ref{sec:ImageAnalysis}. At MTMT, the number of topological defects \textbf{(C)},  the rate of increase in defect number in units of 1/hr \textbf{(D)}, and the defect number normalized by change in the colony area \textbf{(E)}, plotted across the three temperatures.}
\end{figure*}

\begin{figure*}[htbp]
\includegraphics[scale=0.15]{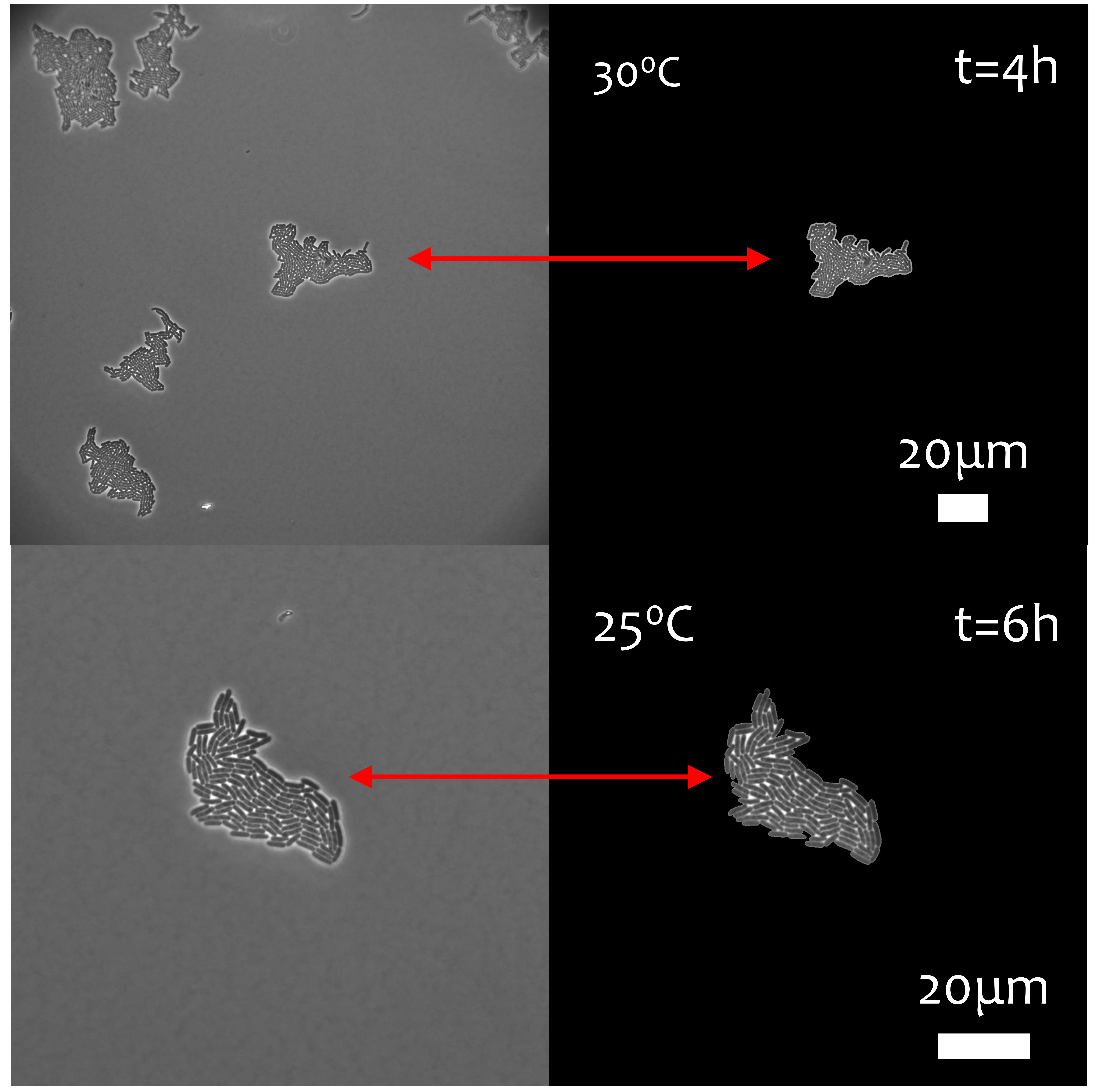}
\caption{\label{fig:suppFig5} \textbf{Single colony extraction for PIV analysis.} A pre-processing is performed prior to the PIV analysis. The two important aims of the analysis are to segment out the background in order to remove any undesired background effects (for instance, particulates), and to detect a single colony to study its growth. Variations in brightness levels can incorporate unwanted noise in the PIV results, thereby rendering the prediction in the local hydrodynamic and transport activity erroneous. In order to reduce errors, single colonies were cropped and analyzed (simultaneous multi colony analysis was not aimed at in the present study). For the purpose of colony extraction, a threshold is applied and area of different labels are extracted from their region properties. Connected area at a particular selected location where a particular point is encompassed (a point chosen on a first cell) is labelled out. This is done across all time steps to extract the bacterial colony. Hereon, the PIV technique is applied on the evolving bacterial colony. The results obtained from PIV pre-processing analysis is given in Supplementary section \ref{sec:Movies} (\textbf{Movie S1} B, E and H for 37\textdegree{}C, 30\textdegree{}C and 25\textdegree{}C, respectively.).}
\end{figure*}

\begin{figure*}[htbp]
\includegraphics[scale=0.14]{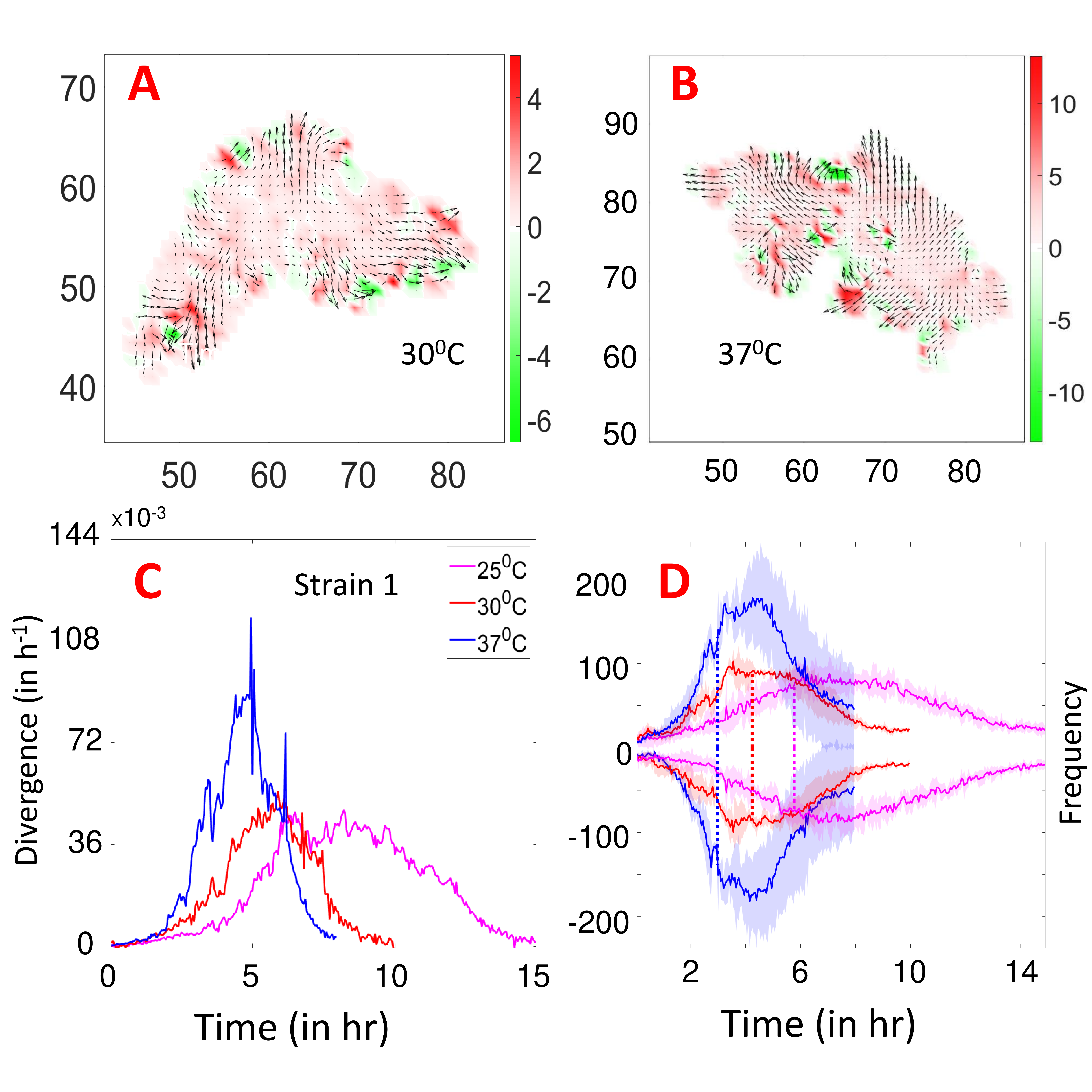}
\caption{\label{fig:suppFig6} \textbf{Divergence and vortex distribution at MTMT.} \textbf{(A)} and \textbf{(B)} The divergence field distribution at 30\textdegree{}C and 37\textdegree{}C, respectively. The divergence strength, as depicted in the colormap, shows strong correlation with the culture temperature. Higher temperature results in a stronger colony and cell growth (Fig S1) and therefore stronger divergence. The colony specification are corresponding to Fig 3A of the main draft. All dimensions are in microns. \textbf{(C)} The mean divergence of a colony is always greater than 0, a signature of an expanding colony. Furthermore, the strength of the divergence decreases with decrease in temperature. Thus quantitatively, the colony activity as a whole regulates the hydrodynamics, both of which can be regulated by the temperature. The trend is same for both the strains across all temperatures, so only a single data is depicted in the figure. \textbf{(D)} Distribution of frequency of positive (anticlockwise) and negative (clockwise) vortices as function of colony age. The figure depicts that the distribution of such vortices are nearly equal across the colony age. This figure is the break-up of Figure \ref{fig:FIG4} presenting the frequency of vortices which gives the total count including both positive and negative vortices. Supplementary section \ref{sec:Movies} \textbf{Movie S1} C, F and I show the time evolution of velocity magnitudes for 37\textdegree{}C, 30\textdegree{}C and 25\textdegree{}C.}
\end{figure*}

\begin{figure*}[htbp]
\includegraphics[scale=0.4]{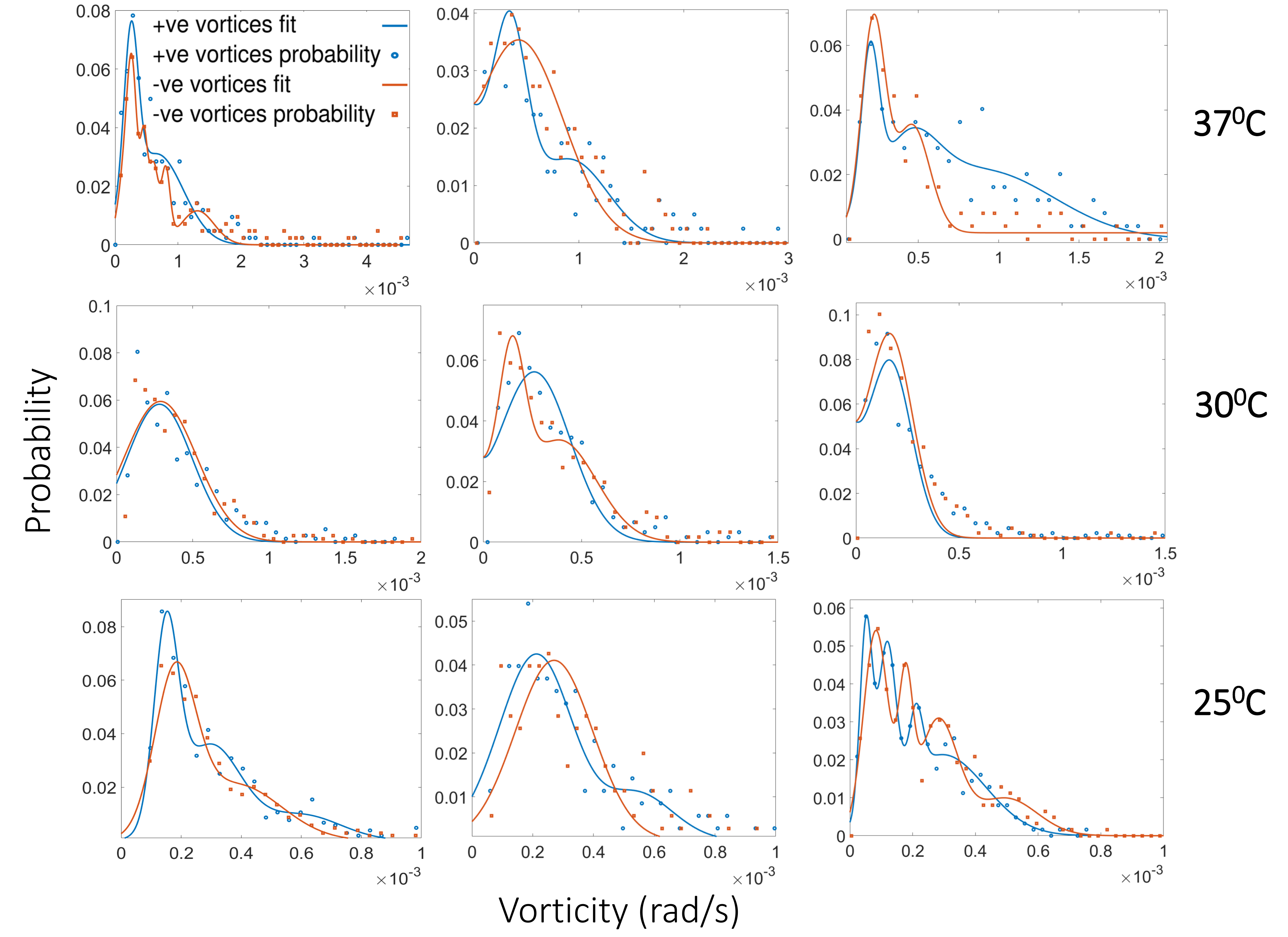}
\caption{\label{fig:suppFig7} \textbf{Log-normal distribution of vortices in a colony at MTMT.} Distribution of phase plots of vorticity strength probability distribution for Strain-1 for each temperature and experimental replicate just before MTMT. A general observation of log-normal distribution of vortex strength is noted.}
\end{figure*}

\begin{figure*}[htbp]
\includegraphics[scale=0.25]{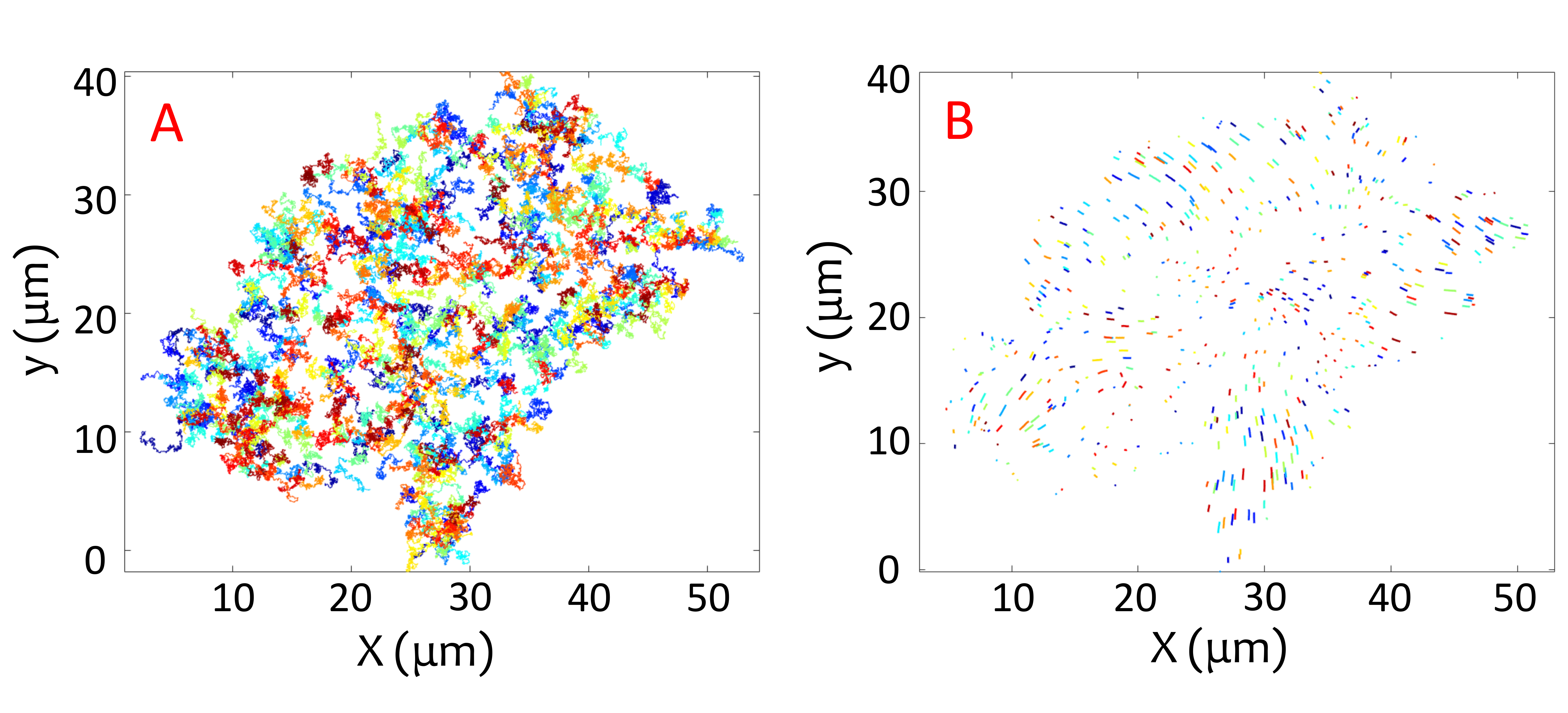}
\caption{\label{fig:suppFig8} \textbf{Simulated tracks of passive particles from the experimental PIV data.} \textbf{(A)} and \textbf{(B)} The particle tracks due to colony expansion corresponding to Figure \ref{fig:Fig5} (A), left and right column respectively. The MSD is derived from these tracks. Panel A shows that the tracks are random which is a signature of strong Brownian motion. Since this corresponds to lower viscosity and particle radius, the Brownian forces are dominant while the fluid momentum transfer from colony motion to the suspended passive particles is suppressed. Panel B shows a more directed motion (also evident from the corresponding MSD-velocity auto-correlation plots in Figure \ref{fig:Fig5}A). For larger particles and higher viscosity, the Brownian motion is outweighed by the momentum generated by the colony expansion.}
\end{figure*}

\begin{figure*}[htbp]
\includegraphics[scale=0.3]{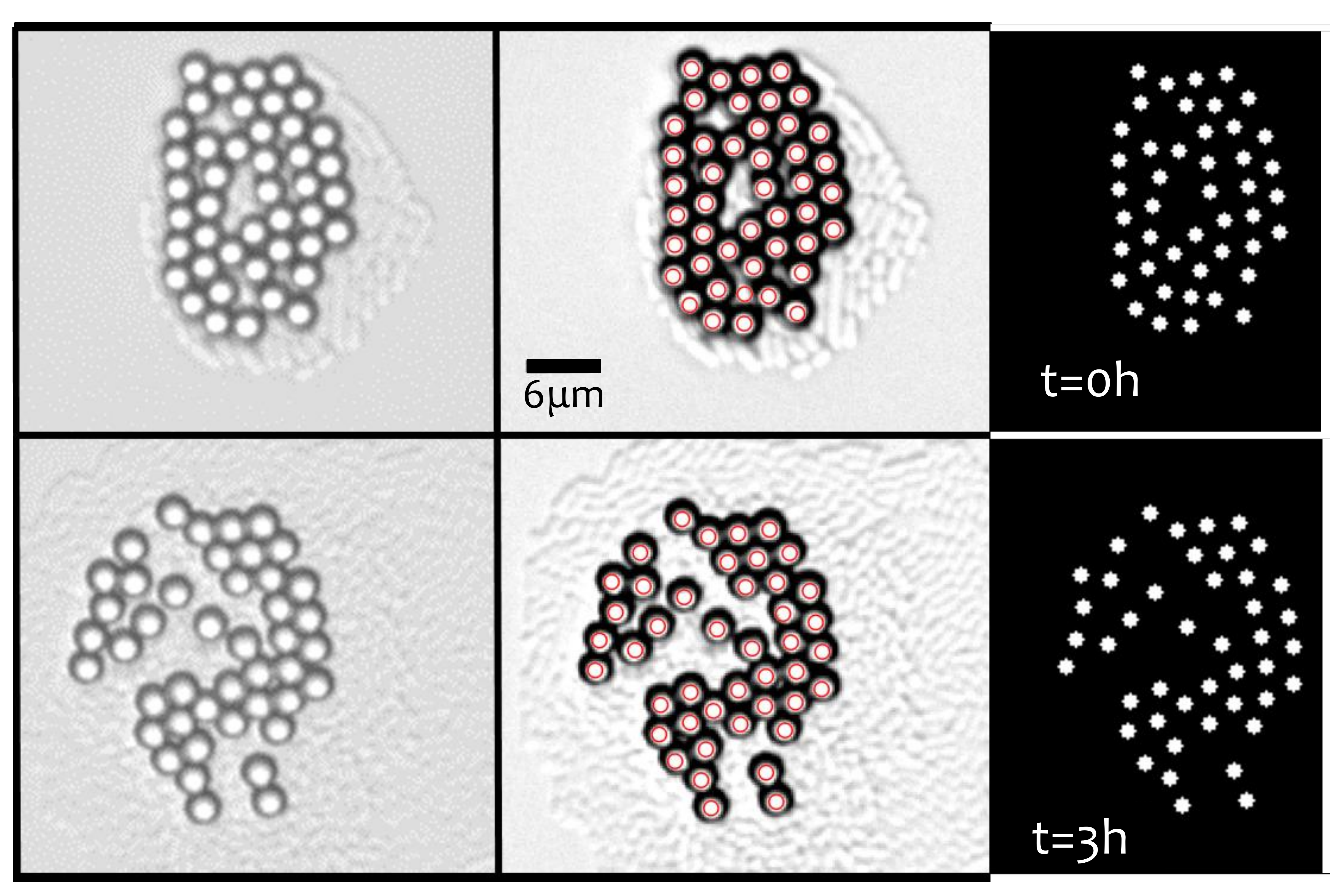}
\caption{\label{fig:suppFig9} \textbf{Identification and trajectory analysis of passive micro-cargo advected by a growing colony.} Experimental visualization of microscale transport induced by a growing bacterial colony at 30\textdegree{}C. The upper row delineates the initial position of the beads and their corresponding identification and segmentation. The beads are identified and extracted using a combination of MATLAB and Ilastik packages. The second panel shows the final position of the beads after 3 hours from the point of initial visualization. A video on the evolution of the bead position for a particular colony (for 37\textdegree{}C and 25\textdegree{}C) is provided in the Supplementary section \ref{sec:Movies} (\textbf{Movie S2}).}
\end{figure*}

\subsection{\label{sec:Movies}Supplementary Movies}
\textbf{(Movie S1)} Videos \textbf{(A)}, \textbf{(D)} and \textbf{(G)} shows the raw image sequence of a growing bacterial colony for three respective temperatures of 370C, 30\textdegree{}C and 25\textdegree{}C. The videos capture one colony for each temperature. Background noise and light reflections can be observed during the capture of the colony growth videos. In order to remove the unwanted noise, the segregation of the bacterial colony must be performed. The extracted single colony evolution with time for each 37\textdegree{}C, 30\textdegree{}C and 25\textdegree{}C is shown in \textbf{(B)}, \textbf{(E)} and \textbf{(H)}, respectively. These extracted images are used as input for the PIV analysis. The video for the evolving velocity field due to the bacterial colony growth is presented in \textbf{(C)}, \textbf{(F)} and \textbf{(I)}, respectively, for 37\textdegree{}C, 30\textdegree{}C and 25\textdegree{}C. From the velocity field magnitude, we observe that at 37\textdegree{}C the colony shows highest hydrodynamic activity reflected by highest velocity magnitude. Furthermore, the velocity field vanishes outside the bacterial colony is due to the initial colony extraction performed on the raw images. 

\textbf{(Movie S2)} Movie shows the evolution of the bead positions and their trajectory for \textbf{(A)} 37\textdegree{}C and \textbf{(B)} 25\textdegree{}C. The trajectories (for each colonies) obtained from such analysis is used to evaluate the MSD and diffusion coefficient of the bead transport. One important aspect which is apparent in the movies is the breakage of the particle agglomeration. Pure Brownian motion, however, remains incapable of breaking the particle agglomeration. Thus, the present phenomena delineates that even for comparatively lower viscosity bacterial motion become biologically significant for agglomerated large molecules. Nevertheless, for non-agglomerating particles, the phenomena will reveal different outcome. We plan for further experiments to learn the effect of inter-particle interaction on bacterial motion-driven transport and understanding the significance of the outcome on biologically relevant situations.

\end{document}